\documentclass{appolb}
\usepackage{amsmath,amssymb,amsfonts} 
\usepackage{graphicx}% Include figure files
\usepackage[numbers,sort&compress]{natbib}
\usepackage{dcolumn}% Align table columns on decimal point
\usepackage{bm}% bold math
\usepackage{nicefrac}
\usepackage{xcolor}
\definecolor{darkblue}{RGB}{0,0,196}
\usepackage[colorlinks=true,linktocpage=true,linkcolor=darkblue,citecolor=red,urlcolor=darkblue]{hyperref}
\usepackage{caption}
\usepackage[caption=false,font=footnotesize,captionskip=-2mm,farskip=-1mm]{subfig}
\usepackage{float}
\usepackage{orcidlink}

\DeclareUnicodeCharacter{03C9}{\ensuremath{\omega}}
\DeclareUnicodeCharacter{039B}{\ensuremath{\Lambda}}

\newcommand{\p}{\partial}
\newcommand{\dd}{\mathrm{d}}

\newcommand{\Ckv}{{\boldsymbol C}_k} 
\newcommand{\Cov}{{\boldsymbol C}_\omega} 
\newcommand{\bvp}{{\boldsymbol b}^\prime} 
\newcommand{\evp}{{\boldsymbol e}^\prime}

\newcommand{\onehalf}{{\nicefrac{1}{2}}} 

\newcommand{\s}{\mathfrak{s}}

\newcommand{\vph}[2]{
  \rule[-#1]{0pt}{#2}
}

\begin{document}

\title{Boost-invariant perfect Fermi--Dirac spin hydrodynamics}
% Force line breaks with \\
%\thanks{A footnote to the article title}}%

 \author{Zbigniew Drogosz
 \thanks{zbigniew.drogosz@alumni.uj.edu.pl}\orcidlink{0000-0001-5133-958X},
 Natalia Łygan\thanks{natalia.lygan@student.uj.edu.pl}\orcidlink{0009-0003-6663-2959}
 \address{Institute of Theoretical Physics, Jagiellonian University, PL-30-348 Krak\'ow, Poland}\\}

\maketitle
\begin{abstract}
We analyze the effect of using the Fermi–Dirac statistics, rather than its Boltzmann approximation, in numerical simulations of perfect spin hydrodynamics of particles with spin 1/2. The system considered is boost invariant, transversely homogeneous, with corrections to the
baryon current and the energy--momentum tensor that are second order in the spin polarization tensor ω, and the spin tensor considered is first order in ω. The study shows the feasibility of this approach, as the special functions defined by integrals that appear in the coefficients in the Fermi--Dirac case can be conveniently parametrized. For sets of initial conditions used in previous works, the differences in parameter evolution between the two underlying particle statistics are about one order of magnitude smaller than corrections coming from spin feedback. We also discuss when and why the numerical solutions of the equations of perfect spin hydrodynamics break down for very large values of spin polarization in one of the geometric configurations considered.
\end{abstract}
%\keywords{spin polarization, relativistic hydrodynamics, heavy-ion collisions}
%Use showkeys class option if keyword

%%%%%%%%%%%%%%%%%%%%%%%%%%%%%%%%%%%%%%%%%%%%%% 
\newpage

\section{Introduction}
\label{sec:introduction}

The search for a suitable theoretical description of spin-polarization phenomena observed in relativistic heavy-ion collisions~\cite{Niida:2024ntm} (in particular the polarization of $\Lambda$ hyperons~\cite{STAR:2017ckg, STAR:2018gyt, STAR:2019erd} and spin alignment of vector mesons~\cite{ALICE:2019aid}) inspired the development of relativistic spin hydrodynamics. This development has followed several paths, including: (I)
determination of the final particle polarization using gradients of hydrodynamic fields on the freezeout hypersurface~\cite{Becattini:2009wh,Becattini:2021iol,Palermo:2024tza}, (II)
kinetic theory approaches~\cite{Florkowski:2017ruc,Shi:2020htn,Hu:2021pwh,  Bhadury:2020puc, Bhadury:2022ulr, Weickgenannt:2019dks, Weickgenannt:2020aaf,  Weickgenannt:2021cuo, Weickgenannt:2022zxs, Weickgenannt:2023nge, Wagner:2024fhf, Banerjee:2024xnd, Bhadury:2024ckc}, (III)
studies founded on the entropy principle~\cite{Li:2020eon,
Hattori:2019lfp, Fukushima:2020ucl,
Biswas:2023qsw, Xie:2023gbo, Daher:2024ixz, Ren:2024pur, Daher:2024bah, 
Fang:2025aig}, (IV) Lagrangian effective hydrodynamics~\cite{Montenegro:2017rbu,Montenegro:2020paq},
(V) formulating spin hydrodynamics as a divergence-type theory~\cite{Abboud:2025shb, Bhadury:2025wuh}.
See Refs.~\cite{Florkowski:2018fap,Becattini:2020ngo,Huang:2024ffg,Florkowski:2024cif} for reviews.

The hybrid approach to spin hydrodynamics, introduced in~\cite{Florkowski:2024bfw, Drogosz:2024gzv}, combines attractive features of several of those pathways. Its description of perfect-fluid spin hydrodynamics is based on kinetic theory of spin-$\onehalf$ particles with classical spin~\cite{Mathisson:1937zz,2010GReGr..42.1011M}, whereas dissipative effects are 
introduced through the entropy-based Israel--Stewart method, thus sidestepping the complexities of the nonlocal collision formalism~\cite{Weickgenannt:2020aaf,  Weickgenannt:2021cuo, Weickgenannt:2022zxs}.
It leads to consistent thermodynamic relations stated in~\cite{Florkowski:2024bfw}.

Although a classical spin description appears to be a contradictory concept and in principle cannot be exact, 
the results were corroborated by 
an approach based on Wigner function with a quantum spin description, which was 
proposed in Ref.~\cite{Bhadury:2025boe}. 
Both in the originally considered Boltzmann approximation and in the Fermi--Dirac case 
\cite{Drogosz:2025ose,Kar:2025qvj}, at the lowest nontrivial order of expansion in the coefficients of the dimensionless spin polarization tensor $\omega_{\alpha \beta} = \Omega_{\alpha \beta}/ T $ (where $\Omega_{\alpha \beta}$ is the spin chemical potential, and $T$ is the temperature), the conserved currents exactly agree between the classical and quantum approaches. The two frameworks differ only in higher-order corrections
(i.e., at least fourth order in coefficients of $\omega_{\alpha \beta}$ in the case of the baryon current $N^\mu$ and the energy--momentum tensor $T^{\mu \nu}$ and at least third order in the case of the spin tensor $S^{\lambda \mu \nu}$), through relative multiplicative factors that have been computed exactly and increase with correction order~\cite{Drogosz:2025iyr}.

It is notable that the applicability range of spin hydrodynamics~\cite{Drogosz:2025ihp}
turns out not to be a constraint in
realistic scenarios, and the nonlinear causality and symmetric hyperbolicity of the equations of motion
are confirmed~\cite{Bhadury:2025boe}. This ensures well-posedness of the initial-value problem and
stability of the theory, thereby implying its usability in numerical simulations of dynamics of spin-polarized fluids, although it does not guarantee that singularities can never form. Properties such as global existence or global regularity of solutions are difficult to prove in theories of relativistic hydrodynamics and require additional assumptions (see, e.g., \cite{Kreiss:1997mk}).

Previous numerical simulations of relativistic hydrodynamics of particles with spin $\onehalf$ were limited to the Boltzmann approximation of the Fermi--Dirac statistics~\cite{Florkowski:2019qdp,Drogosz:2024lkx,Singh:2024cub, Sapna:2025yss}.
This standard simplification is customarily considered to be justified on the assumption of diluteness of the system if the baryon chemical potential $\mu$ is relatively low, and the temperature is high,
rendering the Boltzmann factor small. It is, nevertheless, important to test the numerical effect of this approximation on actual spin hydrodynamics simulations within the approach discussed and with realistic parameter values. 

In this paper, we simulate numerically the equations of perfect-spin hydrodynamics of spin-$\onehalf$ particles following the Fermi--Dirac statistics, with the hydrodynamic tensors expanded up to the second order in the coefficients of $\omega$, and compare the results with the Boltzmann approximation. We use the same geometric setting as Ref.~\cite{Drogosz:2024lkx}, i.e., the Bjorken expansion. That work mentioned the breakdown of solutions for large values of spin polarization only very vaguely. Now we investigate in detail precisely under what conditions the solutions break down in one of the two spin configurations considered, and what the failure mechanism is.

The paper is organized as follows. Section \ref{sec:def} introduces the geometric setting, the coordinate system, and the decomposition of the spin polarization tensor used. Section \ref{sec:hydro} recalls the derivation of the equations of motion of perfect spin hydrodynamics, notes that they are overdetermined in the geometry considered, and describes two configurations with additional symmetry that make the system exactly determined. Section \ref{sec:num} defines the special functions that appear in the Fermi--Dirac case. Section \ref{sec:res} presents the results of numerical simulations. Section \ref{sec:app} investigates the formation of singularities in one of the configurations. Finally, Section \ref{sec:sum} presents the summary and future outlook. Appendix \ref{appendixA} displays the coefficients of the hydrodynamic tensors using the notations and conventions used throughout.

We use the natural units $\hbar = c = k_{\rm B} = 1$, the mostly negative metric tensor $g_{\mu \nu} = {\rm diag}(+1,-1,-1,-1)$, and the convention $\epsilon^{0123} = - \epsilon_{0123} = 1$ for the Levi--Civita symbol.

%%%%%%%%%%%%%%%%%%%%%%%%%%%%%%%%%%%%%%%%%%%%%%
\section{The geometry and the spin polarization tensor}\label{sec:def}

We consider 
the Bjorken expansion, i.e., a boost-invariant, transversely homogeneous geometry \cite{Bjorken:1982qr}, widely used in heavy-ion physics \cite{Florkowski:2010zz}. We use the following orthonormal coordinate basis, whose basis vectors have a form invariant under boosts along the $z$ axis
\begin{align}\begin{split}
U ^\mu &= \frac{1}{\tau}(t,0,0,z)= ( \cosh\eta,0,0, \sinh\eta), \\
X^\mu &= (0,1,0,0), \\
Y^\mu &= (0,0,1,0), \\
Z^\mu &= \frac{1}{\tau}(z,0,0,t)= ( \sinh\eta,0,0,\cosh\eta),\label{eq:basis}
\end{split}\end{align}
where $\tau=\sqrt{t^2 - z^2}$ is the longitudinal proper time, and $\eta= \frac{1}{2} \ln \frac{t+z}{t-z}$ is the spacetime rapidity.
Derivatives transform as
\begin{equation}
\begin{bmatrix}
\partial_t\\ 
\partial_x\\ 
\partial_y\\
\partial_z
\end{bmatrix}
=
\begin{bmatrix}
\cosh\eta&\  0&\ \ 0&  -\frac{1}{\tau} \sinh\eta \\ 
0& \                 1&\ \ 0&  0 \\ 
0& \                 0&\ \ 1&  0  \\
-\sinh\eta&\ 0&\ \ 0& \frac{1}{\tau} \cosh\eta 
\end{bmatrix}
\begin{bmatrix}
\partial_{\tau}\\ 
\partial_x\\ 
\partial_y\\
 \partial_{\eta}
\end{bmatrix}.
\end{equation}
We will denote the $\tau$-derivative with a dot,
\begin{equation}
{\dot f}(\tau) = \frac{\dd f(\tau)}{\dd \tau}.
\end{equation}
The projector $\Delta^{\mu \nu} \equiv g^{\mu \nu} - U^{\mu} U^{\nu}$, orthogonal to the flow in both indices, is expressed in the basis (\ref{eq:basis}) as
\begin{equation}
\Delta^{\mu \nu} = - X^\mu X^\nu - Y^\mu Y^\nu- Z^\mu Z^\nu.
\end{equation}

The antisymmetric rank-2 spin polarization tensor $\omega_{\mu \nu}$ has six degrees of freedom. It admits a decomposition in terms of two four-vectors $k^\mu$ and $\omega^\mu$ orthogonal to the hydrodynamic flow $U^\mu$, 
\begin{equation}
\omega_{\mu \nu}=
k_{\mu} U_{\nu} -k_{\nu} U_{\mu}+\epsilon_{\mu \nu \alpha \beta} U^{\alpha} \omega^{\beta}.
\end{equation}
In the considered geometry, the four-vectors $k^\mu$ and $\omega^\mu$ are
\begin{equation}
k^{\mu} = C_{kx}X^{\mu} + C_{ky}Y^{\mu} + C_{kz}Z^{\mu} = \left( C_{kz}  \sinh \eta, C_{kx}, C_{ky}, C_{kz} \cosh\eta \right),
\end{equation}
\begin{equation}
\omega^{\mu} = C_{\omega x}X^{\mu} + C_{\omega y}Y^{\mu} + C_{\omega z}Z^{\mu} = \left( C_{\omega z}  \sinh\eta, C_{\omega x}, C_{\omega y}, C_{\omega z} \cosh\eta \right),
\end{equation}
where the six coefficients $C_{ki}, \ C_{\omega i}, \ i \in \{x,y,z\}$,
depend on the proper time $\tau$ but not on the transverse coordinates. For brevity, we suppress the argument in $C_{ki}(\tau), \ C_{\omega i}(\tau)$.
Moreover, we adopt a three-vector notation,
\begin{equation}
\Ckv = (C_{kx},C_{ky},C_{kz}), \quad
\Cov = (C_{\omega x},C_{\omega y},C_{\omega z}).
\end{equation}
We will also use the antisymmetric rank-2 tensor $t^{\mu \nu}$ and the four-vector $t^\mu$ defined as
\begin{equation}
t^{\mu\nu} = \epsilon^{\mu \nu \alpha \beta} U_{\alpha} \omega_{\beta},
\end{equation}
\begin{equation}
t^\mu = t^{\mu \nu} k_\nu = \epsilon^{\mu \nu \alpha \beta} k_\nu U_\alpha \omega_\beta.
\end{equation}
It follows that $t^\mu$ is a second-order quantity in the components of the tensor $\omega_{\mu \nu}$.
It can be verified that
\begin{equation}
t^\mu = V_x X^\mu + V_y Y^\mu + V_z Z^\mu = \left(  V_z \sinh\eta, V_x, V_y,  V_z \cosh\eta \right),
\end{equation}
with
\begin{equation}
\boldsymbol V = (V_x, V_y, V_z) = \Ckv \times \Cov.
\end{equation}

\section{Perfect spin hydrodynamics with second-order spin corrections}\label{sec:hydro}

\subsection{Conservation laws}

Perfect spin hydrodynamics is defined by local equilibrium of particles with spin and by conservation laws for the baryon number $N^\mu$, the energy--momentum tensor $T^{\mu \nu}$ and the spin tensor (the spin part of the angular momentum) $S^{\lambda \mu \nu}$,
\begin{align}
\p_\mu N^\mu &= 0,\label{consN}\\
\p_\mu T^{\mu \nu} &= 0,\label{consT}\\
\p_\lambda S^{\lambda \mu \nu} &= 0.\label{consS}
\end{align}
The separate conservation of the spin part of the total angular momentum is justified when the system is characterized by the domination of \mbox{s-wave} scattering~\cite{Coleman:2018mew} and 
long spin relaxation times~\cite{Banerjee:2024xnd} (compare, for example, the systems described in Refs.~\cite{Kapusta:2019sad,Kapusta:2020npk}).
The process of spin--orbit exchange does not appear at the level of perfect fluid and is only introduced 
by dissipative terms, through the nonsymmetric part of the energy--momentum tensor, since the conservation of the total angular momentum \mbox{$\p_\lambda J^{\lambda \mu \nu} = \p_\lambda (x^\mu T^{\lambda \nu} - x^\nu T^{\lambda \mu} + S^{\lambda \mu \nu}) = 0$} implies $\p_\lambda S^{\lambda \mu \nu} = T^{\nu \mu} - T^{\mu \nu}$.

In the considered kinetic-theory framework~\cite{Florkowski:2024bfw}, the conserved currents have the form
\begin{align}
N_{\rm eq}^\mu &=n_u u^{\mu}+n_{t}t^{\mu}, \label{baryontensor}\\
T^{\mu\nu}_{\rm eq} &= \varepsilon u^\mu u^\nu - P_\Delta \Delta^{\mu \nu} \label{enermomtensor}\\ 
&+ (t^\mu u^\nu + t^\nu u^\mu)P_t + (k^\mu k^\nu + \omega^\mu \omega ^\nu) P_{k\omega} ,\nonumber
\\ 
S_{\rm eq}^{\lambda \mu \nu} 
&=u^\lambda \big[ A
\left( k^\mu u^\nu - k^\nu u^\mu \right) + A_1 t^{\mu\nu} \big] \label{spintensor}\\ 
&+ \frac{A}{2} \big[ t^{\lambda \mu} u^\nu - t^{\lambda 
\nu} u^\mu + \Delta^{\lambda \mu} k^\nu - \Delta^{\lambda \nu} k^\mu \big],\nonumber  
\end{align}
where the scalars that multiply vectors and tensors on the right-hand side
are functions of the hydrodynamic variables $T$, $\mu$ (through $\xi \equiv \mu /T$), \mbox{$\omega^2 \equiv \omega_\mu \omega^\mu$} and $k^2 \equiv k_\mu k^\mu$.
See Appendix \ref{appendixA} for the exact expressions.

\begin{figure}[t]
%\subfloat[][\centering]
{\includegraphics[width=0.5\textwidth]{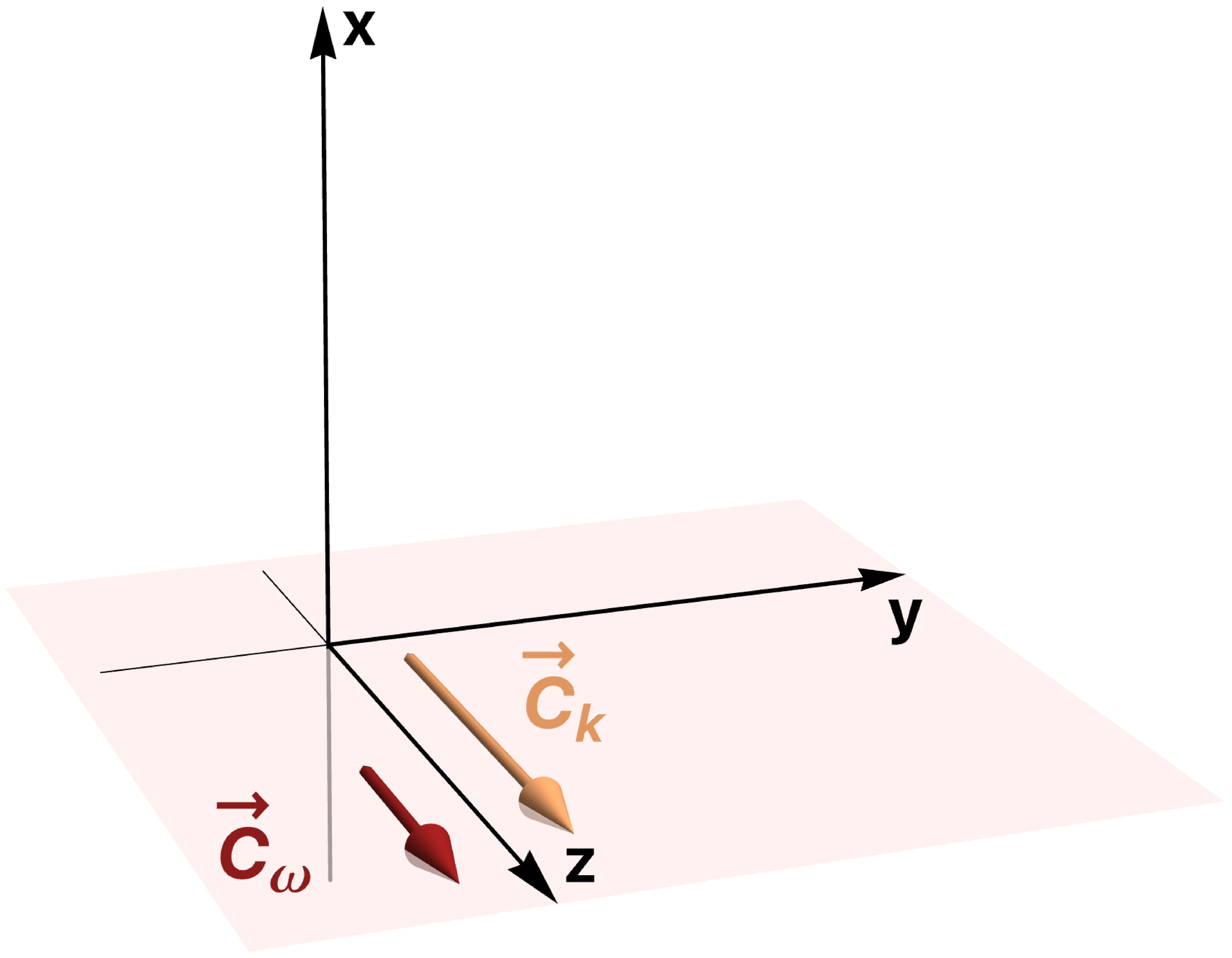}}
\quad
%\subfloat[][\centering]
{\includegraphics[width=0.4\textwidth]{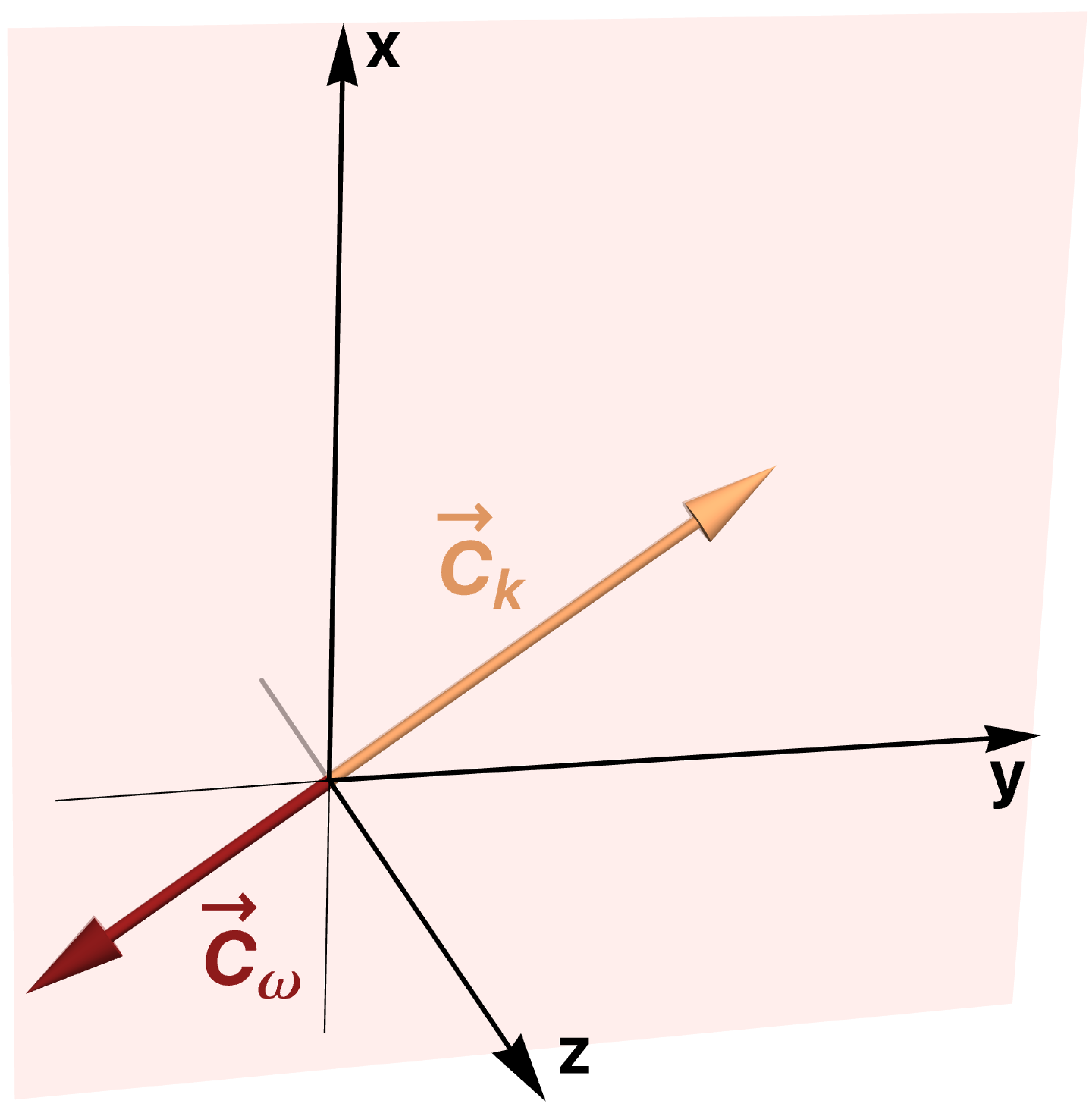}}
\caption{Schematic view of the longitudinal and the transverse configuration (left and right panel respectively). }\label{fig:sch}
\end{figure}

Substituting the forms of the tensors \eqref{baryontensor}--\eqref{spintensor} into the equations \eqref{consN}--\eqref{consS} leads to 
\begin{equation}\label{eq:c1}
\frac{\dd n_u}{\dd \tau} +\frac{n_u}{\tau} = 0,
\end{equation}
\begin{align}\begin{split}
\label{eq:c2} \partial_{\mu} T^{\mu \nu} &= \Bigg[ \dot\varepsilon + \frac{\varepsilon + P_\Delta}{\tau} + \frac{P_{k \omega}}{\tau} ( C^2_{kz} + C^2_{\omega z} ) \Bigg] U^\nu  \\
&+ \Bigg[ \big( \dot{P_t} + \frac{P_t}{\tau} \big) V_x + P_t \dot{V_x} \Bigg] X^\nu + \Bigg[ \big( \dot{P_t} + \frac{P_t}{\tau} \big) V_y + P_t \dot{V_y} \Bigg] Y^\nu \\
&+ \Bigg[ \big( \dot{P_t} + \frac{P_t}{\tau} \big) V_z+
P_t \big( \dot{V_z} + \frac{V_z}{\tau} \big)  
\Bigg] Z^\nu = 0, 
\end{split}\end{align}
\begin{align}\begin{split}
\label{eq:c3}\p_\lambda S^{\lambda \mu \nu} &= \frac{1}{\tau}
\left[ A \left( k^\mu U^\nu - k^\nu U^\mu \right) +A_1 t^{\mu \nu} \right] + U^\nu \p_\tau( A k^\mu 
)- U^\mu \p_\tau( A k^\nu ) \\
&+ \dot{A}_1 t^{\mu \nu} + A_1 \dot{t}^{\mu \nu} 
+ \frac{1}{2}  \left( t^{\lambda \mu} U^\nu - t^{\lambda \nu} U^\mu \right) \p_\lambda A \\ 
&+ \frac{A}{2} \Big[  U^\nu \p_\lambda t^{\lambda \mu}+ t^{\lambda \mu} \p_\lambda U^\nu - U^\mu \p_\lambda t^{\lambda \nu} - t^{\lambda \nu} \p_\lambda U^\mu \\ 
&- \frac{U^\mu}{\tau} k^\nu + \Delta^{\lambda \mu} \p_\lambda k^\nu + \frac{U^\nu}{\tau} k^\mu - \Delta^{\lambda \nu} \p_\lambda k^\mu \Big] = 0.
\end{split}\end{align}
Vanishing of every independent component of these divergences is equivalent to eleven ordinary differential equations for eight unknown functions of the proper time, $T(\tau)$, $\xi(\tau)$, $\Cov(\tau)$, $\Ckv(\tau)$,
\begin{align}
\dot n_u +\frac{n_u}{\tau} &= 0,\label{eq:syst1}\\
\dot\varepsilon + \frac{\varepsilon + P_\Delta}{\tau} + \frac{P_{k \omega}}{\tau} \left( C^2_{kz} + C^2_{\omega z} \right) &= 0,\label{eq:syst2}\\
\big( \dot{P_t} + \frac{P_t}{\tau} \big) V_i + P_t \dot{V}_i + P_t \delta_{iz} \frac{V_i}{\tau} &= 0\label{eq:syst3}\\
A \dot C_{ki} + Q_i C_{ki} &= 0,\label{eq:syst4}\\
A_1 \dot C_{\omega i} + R_i C_{\omega i} &= 0,\label{eq:syst5}
\end{align}
where the index $i$ takes the values $x,y,z$, and $\delta_{ij}$ is the Kronecker delta. The quantities $Q_i$ and $R_i$ are defined as
\begin{align}\begin{split}
Q_x = Q_y &= \dot A + \frac{3A}{2\tau} , \quad Q_z = \dot A + \frac{A}{\tau},\\
R_x = R_y &= \dot A_1 + \frac{2 A_1 - A}{2\tau}, \quad R_z = \dot A_1 + \frac{A_1}{\tau}.
\end{split}\end{align}
Without further constraints, the system of equations (\ref{eq:syst1})--(\ref{eq:syst5}) is overdetermined. However, there exist special configurations where $\Ckv$ and $\Cov$ obey additional symmetries that result in an exactly determined system. We discuss two such configurations, a longitudinal one and a transverse one (see Fig.~\ref{fig:sch} for a schematic view). Their geometries are the same in both the Fermi--Dirac and the Boltzmann cases, but the coefficients are expressed by different functions, which leads to quantitatively different solutions.

\subsection{Configurations with additional symmetries}

One spin configuration for which the equation count of the system \eqref{eq:syst1}--\eqref{eq:syst5} is reduced and equal to the number of unknowns is a longitudinal configuration in which $\Ckv$ and $\Cov$ point along the $z$ axis,
\begin{equation}
\Ckv = (0,0,C_{kz}), \quad \Cov=(0,0,C_{\omega z}), \quad \boldsymbol V = (0,0,0).
\end{equation}
This constraint produces a system of four independent equations for four unknown functions $T(\tau)$, $\mu(\tau)$, $C_{kz}(\tau)$, $C_{\omega z}(\tau)$,
\begin{align}\label{eq:l1}
\dot{n}_u  &= - \frac{n_u}{\tau}, 
  \\
\dot{\varepsilon} &= - \frac{\varepsilon + P_\Delta}{\tau} - \frac{P_{k \omega}}{\tau} \left( C^2_{kz} + C^2_{\omega z} \right), \label{eq:l2}\\
\dot{C}_{k z} &= - \bigg(\frac{\dot A}{A} + \frac{1}{\tau} \bigg) \, C_{k z}, \label{eq:l3}\\
\dot{C}_{\omega z} &= - \bigg( \frac{\dot A_1}{A_1} + \frac{1}{\tau} \bigg) \, C_{\omega z},\label{eq:l4}
\end{align}
where the coefficients $n_u$, $\varepsilon$, $P_\Delta$ are functions of $(T,\mu,k^2, \omega^2)$, i.e., \linebreak\mbox{$(T,\mu,-C_{kz}^2, -C_{\omega z}^2)$}, and the coefficients $P_{k \omega}$, $A$, $A_1$ are functions of $(T,\mu)$.
Spin feedback enters the baryon number and energy equations both through the last two arguments of those coefficients and through the term with $(C_{kz}^2 + C_{\omega z}^2)$ in (\ref{eq:l2}).

Another geometry for which the number of equations is equal to the number of unknowns is a transverse configuration in which $\Ckv$ and $\Cov$ lie in the x-y plane,
\begin{equation}
\Ckv = \left( C_{kx}, C_{ky}, 0 \right), \quad \Cov = \left( C_{\omega x}, C_{\omega y}, 0 \right), \quad \Ckv\parallel \Cov, \quad \boldsymbol V = 0,
\end{equation}
This choice leads to six differential equations for six unknown functions,
\begin{align}\label{eq:t1}
\dot n_u  &= - \frac{n_u}{\tau},\\
\dot\varepsilon &= - \frac{\varepsilon + P_\Delta}{\tau},\label{eq:t2}   \\
A \dot C_{k x} &= - \Big( \dot A + \frac{3A}{2\tau} \Big) \, C_{k x}\\
A \dot C_{k y} &= - \Big( \dot A + \frac{3A}{2\tau}\Big) \, C_{k y},\label{eq:t3} \\ 
A_1 \dot C_{\omega x} &= - \Big(\dot A_1 + \frac{2 A_1 - A}{2\tau}\Big) \, C_{\omega x}, \\
A_1 \dot C_{\omega y} &= - \Big(\dot A_1 + \frac{2 A_1 - A}{2\tau}\Big) \, C_{\omega y}.\label{eq:t4},
\end{align}
where the coefficients $n_u$, $\varepsilon$, $P_\Delta$ are functions of $(T,\mu,k^2, \omega^2)$, i.e., \linebreak\mbox{$(T,\mu,-C_{kx}^2-C_{ky}^2, -C_{\omega x}^2-C_{\omega y}^2)$}, and the coefficients $P_{k \omega}$, $A$, $A_1$ are functions of $(T,\mu)$. In this case, spin feedback enters the first two equations through dependence of $n_u$, $\varepsilon$ and $P_\Delta$ on $k^2$ and $\omega^2$.

\section{Special functions}\label{sec:num}

\begin{figure}[t]
\includegraphics[width=\linewidth]{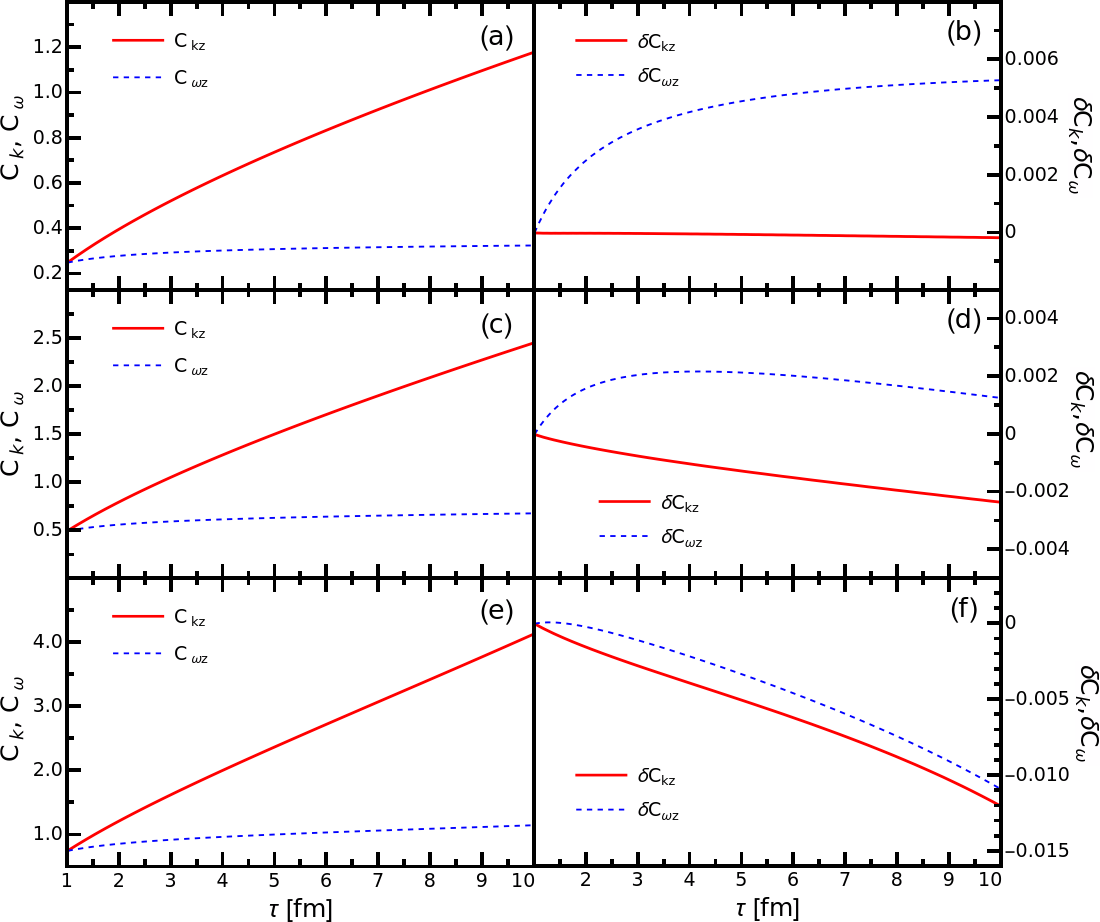}
\caption{Proper-time dependence of coefficients $C_{kz}$, $C_{\omega z}$ of the spin polarization tensor in the longitudinal configuration in the Fermi--Dirac case (left column) and the relative differences between the Fermi--Dirac and the Boltzmann case, defined as $\delta C_{kz} \equiv C_{kz {\rm FD}}/C_{kz {\rm B}} - 1$, $\delta C_{\omega z} \equiv C_{\omega z {\rm FD}}/C_{\omega z {\rm B}} - 1$ (right column). The initial values of coefficients $C_{kz}$ and $C_{\omega z}$ are $0.25$ (top row), $0.5$ (middle row), or $0.75$ (bottom row).}\label{fig:l}
\end{figure}

\begin{figure}[t]
\includegraphics[width=\linewidth]{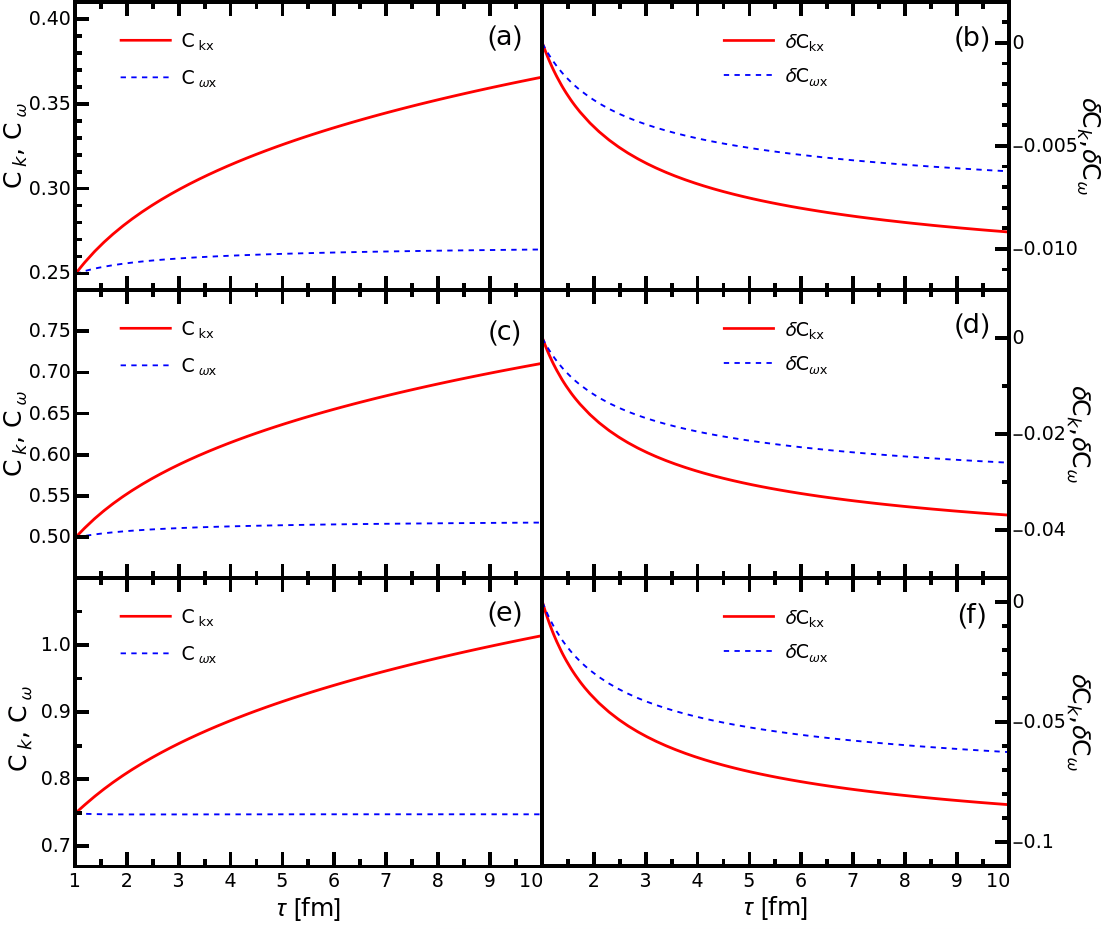}
\caption{Proper-time dependence of coefficients $C_{kx}$, $C_{\omega x}$, of the spin polarization tensor in the transverse configuration in the Fermi--Dirac case (left column) and the relative differences between the Fermi--Dirac and the Boltzmann case, $\delta C_{kx} \equiv C_{kx {\rm FD}}/C_{kx {\rm B}} - 1$, $\delta C_{\omega x} \equiv C_{\omega x {\rm FD}}/C_{\omega x {\rm B}} - 1$ (right column). The initial values of coefficients $C_{kx}$, $C_{ky}$, $C_{\omega x }$, $C_{\omega y}$ are $0.25$ (top row), $0.5$ (middle row), or $0.75$ (bottom row). Owing to the symmetry of the system, the $y$-direction coefficients remain equal to the $x$-direction coefficients throughout the simulation.}\label{fig:t}
\end{figure}

The coefficients of the conserved current tensors  depend on thermodynamic variables through the functions $J_{mn0}(\xi,z)$, where $m,n = 0,1,2,\dots$, and their first two derivatives with respect to $\xi$, denoted as $J_{mn1}(\xi,z)$ and $J_{mn2}(\xi,z)$ (see Appendix \ref{appendixA}). They are defined by integrals
\begin{align}\label{eq:js}
J_{mn0}(\xi,z) &\equiv \int_0^\infty \frac{\sinh^m y \cosh^n y}{\exp(-\xi + z \cosh y) + 1} \dd y,\\
J_{mn1}(\xi,z) &\equiv \frac{\p}{\p\xi} J_{mn0}(\xi,z) = \int_0^\infty \frac{\sinh^m y \cosh^n y}{2 + 2\cosh(\xi - z \cosh y)} \dd y,\label{eq:js1}\\
J_{mn2}(\xi,z) &\equiv \frac{\p^2}{\p\xi^2} J_{mn0}(\xi,z) =\nonumber\\
&=\int_0^\infty \frac{\sinh^m y \cosh^n y \sinh(\xi-z \cosh y)}{2 \big(1 + \cosh(\xi - z \cosh y) \big)^2} \dd y.\label{eq:js2}
\end{align}
Unlike the Bessel function of the second kind $K_n(z)$, which appears in the coefficients in the Boltzmann approximation, these functions are not included in standard numerical packages.
In order to use them effectively in differential equations, we constructed interpolating functions.
Since the asymptotic behavior of functions $J_{mnk}(\xi,z)$, for large $\xi$ or $z$ is exponential, we tabulated values of the logarithm of each of the required functions (14 functions in total).
To ensure coverage of the whole relevant parameter region, we tabulated the logarithms of $J_{mnk}(\xi,z)$, with $\xi$ ranging from $-100$ to $100$ with step $0.5$, and $z$ ranging from $0.1$ to $100$ with step $0.1$. The numerical values of partial derivatives of logarithms of $J_{mnk}(\xi,z)$, with respect to $\xi$ and $z$, were also tabulated on the same grid and supplied to Wolfram Mathematica's \texttt{Interpolation} function to be used as constraint on the interpolant.

%%%%%%%%%%%%%%%%%%%%%%%%%%%%%%%%%%%%%%%%%%%%%

\section{Results}\label{sec:res}
Figures~\ref{fig:l} and \ref{fig:t} present the results of solving numerically the differential equation systems \eqref{eq:l1}--\eqref{eq:l4} and \eqref{eq:t1}--\eqref{eq:t4}, respectively.
The simulations were conducted with the following parameter choice: the initial proper time \mbox{$\tau_0 = 1 \ {\rm fm}$}, the final proper time \mbox{$\tau_f = 10 \ {\rm fm}$}, the particle mass equal to that of the $\Lambda$ hyperon, \mbox{$m= 1.116 \ {\rm GeV}$}, the initial temperature and baryon chemical potential having values typical for moderate-energy heavy-ion collisions, \mbox{$T_0 = 155 \ {\rm MeV}$}, \mbox{$\mu_0 = 800 \ {\rm MeV}$}.
As we are mostly interested in the evolution of spin degrees of freedom, we plot components of $\Ckv$ and $\Cov$ only (the impact of spin feedback on the evolution of temperature $T$ and baryon chemical potential $\mu$ was demonstrated to be small in a previous study \cite{Drogosz:2024lkx}).
The top row in both figures corresponds to the initial values of the coefficients $C_i$ chosen as $0.25$, the middle row $0.5$, and the bottom row $0.75$. 

In both configurations, the values of the coefficients $\Ckv$ grow monotonically, and their increase is faster for larger initial values. On the other hand, the coefficients $\Cov$ either grow slowly or even slowly decrease (Fig.~\ref{fig:t}, panel (e)). The right column in each figure shows the relative difference between the Fermi--Dirac and the Boltzmann case, defined as \mbox{$\delta C_{i} \equiv C_{i \rm{FD}}/C_{i \rm B} - 1$}.

For the choices of parameters shown in the figures, the relative differences between the Fermi--Dirac case and its Boltzmann approximation are small, though not always negligible, reaching about $0.085$ in one case, Fig.~\ref{fig:t}, panel (f).

For some even larger initial coefficient values (not shown in the figures), we observed formation of very steep gradients and loss of convergence in the longitudinal configuration, with little difference between the Boltzmann and Fermi--Dirac cases. On the other hand, in the transverse configuration, the solver reached $\tau_f$ for any initial coefficient values tested. This is explored in more detail in the next section.

\section{Applicability of the theory expanded to the second order in $\omega$}\label{sec:app}

It was observed that the numerical solutions in the longitudinal case develop a singularity (a finite-time blow-up) 
before $\tau = 10 \ {\rm fm}$ for initial magnitudes of coefficients of the spin polarization tensor $\omega_{\mu \nu}$ above approximately $0.87$ (with values of $m$, $T_0$ and $\mu_0$ as given in the previous section). Remarkably, simulations of the transverse case
were free from singularities for any initial values of coefficients of $\omega_{\mu \nu}$, even of the order of $10^6$.

It is worth analyzing this in light of previous results on applicability, stability and causality of perfect spin hydrodynamics.

\subsection{Applicability}

Perfect spin hydrodynamics with classical description of spin is applicable if the integral appearing in the generating function
\begin{equation}\label{ncl}
n_{\rm cl} =2\int \dd P   \cosh \xi\exp\left(- p \cdot \beta \right)
\int \dd S \exp\left(\frac{1}{2} \omega : s\right)
\end{equation}
converges.
This leads to a criterion that limits allowable values of the spin polarization tensor,
\begin{equation}\label{criterion}
\sqrt{ {\bvp}^2 + {\evp}^2 + 2 |\evp \times \bvp|} < \frac{m}{T\s}
\end{equation}
(derived in \cite{Drogosz:2025ihp} for particles following the Boltzmann statistics but valid for the Fermi--Dirac case as well, as mentioned in \cite{Kar:2025qvj}).
Primes denote quantities in the local rest frame (LRF) of the fluid element, $\s= \sqrt{3/4}$ is the normalization of the spin four-vector, $m$ denotes particle mass, $T$ denotes temperature, while
\mbox{$\boldsymbol e = (e^1,e^2,e^3)$} and $\boldsymbol b = (b^1,b^2,b^3)$ are electriclike and magneticlike three-vectors through which the spin polarization tensor may be expressed as~\cite{Florkowski:2017dyn}
\begin{eqnarray}
\omega_{\mu\nu} = 
\begin{bmatrix}
0     &  e^1 & e^2 & e^3 \\
-e^1  &  0    & -b^3 & b^2 \\
-e^2  &  b^3 & 0 & -b^1 \\
-e^3  & -b^2 & b^1 & 0
\end{bmatrix}.
\end{eqnarray}
If the generating function \eqref{ncl} converges, then the tensors
\begin{align}
N_{\rm eq}^\mu &= - \frac{\p^2 n_{\rm cl}}{\p \beta \p \xi},\\
T^{\mu \nu}_{\rm eq} &= \frac{\p^2 n_{\rm cl}}{\p \beta_\mu \p\beta_\nu},\\
S_{\rm eq}^{\lambda \mu\nu} &= -\frac{\p^2 n_{\rm cl}}{\p \beta_\lambda \p\omega_{\mu \nu}}
\end{align}
are finite as well. Conversely, finite energy density implies convergence of \eqref{ncl} and thus the satisfaction of \eqref{criterion}.

It is interesting to note that the expression \eqref{criterion} takes a particularly simple form in the Bjorken expansion.
Below we show that in such a geometry the squared lengths of three-vectors $\boldsymbol e$ and $\boldsymbol b$ as well as of their vector product correspond to squared magnitudes of four-vectors $k$, $\omega$ and $t$.
In the LRF,
\begin{equation}
\omega'_{\mu \nu} = k'_\mu \delta_{\nu 0} - k'_\nu \delta_{\mu 0} + \epsilon_{\mu \nu 0 \beta} \ \omega'^\beta,
\end{equation}
from which we can read that $k'_\mu = (0,-\evp)$, $\omega'_\mu = (0,-\bvp)$, and thus \mbox{$-k^\mu k_\mu = \evp{}^2$}, $-\omega^\mu \omega_\mu = \bvp{}^2$, and $- k^\mu \omega_\mu = \evp \cdot \bvp$. Using these results,
\begin{align}\begin{split}
t^\mu t_\mu &= \epsilon^{\mu \nu \alpha \beta} k_\nu U_\alpha \omega_\beta \, \epsilon_{\mu \rho \sigma \tau} k^\rho U^\sigma \omega^\tau = \epsilon^{\mu \nu 0 \beta} k'_\nu \omega'_\beta \, \epsilon_{\mu \rho 0 \tau}k'^\rho \omega'^\tau \\
&= \epsilon^{0 \mu \nu \beta} k'_\nu \omega'_\beta \, \epsilon_{0 \mu \rho \tau}k'^\rho \omega'^\tau = -|\evp \times \bvp|^2
\end{split}\end{align}
Therefore, the expression \eqref{criterion} takes the form
\begin{equation}
\sqrt{- k^\mu k_\mu - \omega^\mu \omega_\mu + 2 \sqrt{- t^\mu t_\mu}} < \frac{m}{T\s}.
\end{equation}
Subsequently, $k$, $\omega$, and $t$ can be expressed in terms of $\Ckv$ and $\Cov$,
\begin{equation}\label{criterionBI}
\sqrt{\Ckv^2(\tau) + \Cov^2(\tau) + 2 \left| \Ckv(\tau) \times \Cov(\tau)\right|} < \frac{m}{T(\tau)\s}.
\end{equation}

In fact, however, the criteria \eqref{criterion} and \eqref{criterionBI} are irrelevant for the theory expanded to any finite order in $\omega$ because the exponential function of $\frac12 \omega : s$ appearing in (\ref{ncl}) is replaced by a polynomial, making the resulting integrals always convergent, e.g.,
\begin{equation}
n_{\rm cl} =2\int \dd P   \cosh \xi\exp\left(- p \cdot \beta \right)
\int \dd S \left(1 +\frac{1}{2} \omega : s + \frac{1}{8} (\omega : s)^2 \right)
\end{equation}
in the second-order expansion. Therefore, this is not the mechanism of singularity creation that we observed in the numerical experiments.

\subsection{Failure mode of the longitudinal configuration equation system}

In the longitudinal configuration, three of the four equations \eqref{eq:l1}--\eqref{eq:l4} can be integrated,
\begin{align}
\frac{\dd}{\dd \tau} \left( \tau n_u\right) = 0,\label{inv1}\\
\frac{\dd}{\dd \tau} \left( \tau C_{k z} A \right) = 0, \label{inv2}\\
\frac{\dd}{\dd \tau} \left( \tau C_{\omega z} A_1 \right) = 0.\label{inv3}
\end{align}
Hence,
\begin{align}\label{eq:ckzcoz}
C_{k z}(\tau) = \frac{C_{kz,0} A_0 \tau_0}{\tau A(T(\tau), \mu(\tau))}, \qquad C_{\omega z}(\tau) = \frac{C_{\omega z,0} A_{1,0} \tau_0}{\tau A_1(T(\tau), \mu(\tau))}.
\end{align}

Let us denote with a tilde a reduced version of any coefficient, obtained by substituting $C_{kz}(\tau)$ and $C_{k \omega}(\tau)$ as in Eq.~\eqref{eq:ckzcoz}. (This introduces explicit dependence on $\tau$.)
Then the first two equations become a $2\times 2$ linear system for
$\dot T$ and $\dot\mu$,
\begin{align}
0 &= \tau \frac{\dd}{\dd\tau}\widetilde n_u + \widetilde n_u, \label{eq:nc}\\
0 &= \tau \frac{\dd}{\dd\tau}\widetilde\varepsilon + \widetilde\varepsilon + \widetilde P_\Delta + \left( C_{kz}(\tau)^2 + C_{\omega z}(\tau)^2 \right)\,\widetilde P_{k \omega}.
\end{align}
Equation~\eqref{eq:nc} is integrable,
\begin{equation}
\tau n_u(\tau)=N_n= \tau_0 n_u\left(T(\tau_0), \mu(\tau_0),C_{kz}(\tau_0), C_{\omega z}(\tau_0) \right),
\end{equation}
and lets us define an algebraic constraint for $(T,\mu)$ at each $\tau$,
\begin{equation}
g(T,\mu,\tau) = \tau\,\widetilde n_u(T,\mu,\tau) - N_n = 0.
\label{eq:g_constraint}
\end{equation}
Now let us differentiate $g(T,\mu,\tau)=0$,
\begin{equation}
0 = g_T \dot T + g_\mu \dot\mu + g_\tau,
\end{equation}
obtain $\dot\mu$,
\begin{equation}
\dot\mu = -\frac{g_T \dot T + g_\tau}{g_\mu},
\label{eq:mu_dot_from_g}
\end{equation}
and write the energy equation in the form
\begin{equation}
0 = \tau\left(\widetilde\varepsilon_T \dot T + \widetilde\varepsilon_\mu \dot\mu
+\widetilde\varepsilon_\tau\right)
+\widetilde\varepsilon + \widetilde P_\Delta
+ \left( C_{kz}(\tau)^2
+ C_{\omega z}(\tau)^2 \right)\widetilde P_{k \omega}
\label{eq:energy_linear}
\end{equation}
Substituting \eqref{eq:mu_dot_from_g} into \eqref{eq:energy_linear} reduces the equation system to a single explicit ordinary differential equation for $T$,
\begin{equation}
\dot T
=
-\frac{
\tau\,\widetilde\varepsilon_\tau
+\widetilde\varepsilon + \widetilde P_\Delta
+ \left(C_{kz}(\tau)^2
+ C_{\omega z}(\tau)^2 \right) \widetilde P_{k \omega}
- \tau\,\widetilde\varepsilon_\mu\,\dfrac{g_\tau}{g_\mu}
}{
\tau\left(\widetilde\varepsilon_T
- \widetilde\varepsilon_\mu\,\dfrac{g_T}{g_\mu}\right)
},
\label{eq:T_dot_single}
\end{equation}
whereas $\dot\mu$ follows from \eqref{eq:mu_dot_from_g}.
This form shows clearly a possible failure mode: the denominator of \eqref{eq:T_dot_single} becoming zero.

And, indeed, numerical computations confirm that for the $T_0$, $\mu_0$, and $\tau_0$ considered, if one probes increasing values of $C_0 \equiv C_{kz0} = C_{\omega z 0}$,
the denominator of the expression above goes to zero and changes sign at \mbox{$C_0 \approx 2.0568$} (value for the Boltzmann case).
In simulations, for choices of $C_0$ near this value, 
the blow-up occurs virtually immediately, at the first time steps. As we consider longer time scales, larger intervals of initial conditions for $C_0$ cause the solution to reach eventually the point of vanishing denominator, and with $\tau_0=1 \,\rm{fm}$ and $\tau_f = 10 \, \rm{fm}$, configurations with starting conditions in the interval approximately $0.8697<C_{0} <4.4809$ exhibit a singularity.

\subsection{Analysis of the transverse configuration equation system}

Although this subsection does not purport to offer a proof of global existence of solutions in the transverse case, it presents some 
intuitive arguments why the equations of the transverse configuration behave better than those of the longitudinal one.

Relations corresponding to Eqs.~\eqref{inv2} and \eqref{inv3} are
\begin{equation}\label{invt}
\frac{\dd}{\dd\tau}\bigl(\tau ^{3/2}C_{k i}A\bigr) = 0
\end{equation}
and
\begin{equation}\label{eq:Cw_total_derivative}
\frac{\dd}{\dd\tau}\bigl(\tau\,C_{\omega i}A_1\bigr) = \frac{A}{2}C_{\omega i},
\end{equation}
with $i \in \{x,y\}$.
Using $W_i \equiv \tau\,C_{\omega i}A_1$, the equation above is equivalent to
\begin{equation}\label{eq:Wi_ode}
\dot W_i = \frac{A(T,\mu)}{2 \tau\,A_1(T,\mu)}\,W_i.
\end{equation}
This equation does not admit an analytical solution, and thus the transverse configuration equation system cannot be reduced to a single differential equation. However, one may observe that since $A_1>0$ and $A<0$ (both proportional to positive Bessel functions, with a
relative minus sign), the prefactor in \eqref{eq:Wi_ode} is negative, and $W_i$ is damped compared to \eqref{inv3}, which in this notation is simply $W_z = \operatorname{const.}$
Additionally, in \eqref{invt} the term $C_{ki} A$ is counterbalanced by a higher power of $\tau$ than in \eqref{inv1}.

In effect, the equation system of the transverse configuration appears to have solutions in the considered time frame for any tested initial values of $C_{ki}$ and $C_{\omega i}$, even of the order of $O(10^6)$.

\subsection{Causality and stability vs.~global existence of solutions}

Perfect spin hydrodynamics can be cast into the divergence-type theory (DTT) framework, as described in \cite{Bhadury:2025wuh}, and the criteria of causality and stability are satisfied. This holds as well for the theory expanded to a finite order in $\omega$, as can be easily verified. Nevertheless, these properties do not imply global existence or global regularity of solutions.

Global existence of solutions in DTTs demands additional assumptions; see, e.g., the paper \cite{Kreiss:1997mk}, which requires dissipation (with additional conditions that cause a theory to be what the authors call totally dissipative) and the compact spatial topology of a flat 3-torus.

Recent works describe shocks in numerical simulations of DTTs for conformal fluids \cite{Montes:2023dex} and
finite-time formation of steep gradients in Israel-Stewart theories \cite{Bemfica:2025gws}.

Formation of singularities in finite time from smooth initial data has been observed in numerical simulations of viscous BDNK (Bemfica, Disconzi, Noronha, Kovtun) hydrodynamics as well, regardless of its being a causal, stable, strongly hyperbolic and locally well-posed theory \cite{Keeble:2025bkc}.

Thus, it appears that formation of singularities is a generic phenomenon in numerical simulations of relativistic hydrodynamics. Seen from this viewpoint, it is remarkable that the numerical simulations of the second-order theory in the transverse configuration never developed singularities within the considered proper-time range, and the solver reached $\tau_f$ starting from any initial $C_k$ and $C_\omega$.

\section{Summary and outlook}\label{sec:sum}

In this work, the numerical
study of perfect spin hydrodynamics of \mbox{spin-$\onehalf$} particles,
with corrections that are second order in the components of the spin polarization tensor, was refined by
applying expressions 
derived for particles obeying the Fermi--Dirac statistics rather than its Boltzmann approximation. 

We have found that using the exact Fermi--Dirac statistics instead of its Boltzmann approximation
did not change the general behavior of the parameters $C_{k}$ in the simulations (i.e., monotonic increase).
However, it led to relative differences in spin polarization of up to about $8.5 \%$, which is not negligible. Presumably, making the system less dilute and increasing the initial baryon chemical potential $\mu$ would increase the relative differences further.
The relative differences were the highest in simulations with the highest initial values of the coefficients of the spin polarization tensor $\omega_{\mu \nu}$, and those values were probably higher than those typically occurring in heavy-ion collisions.

Tabulating and interpolating the special functions that appear in the coefficient formulas in the Fermi--Dirac case involves numerical integration and is computationally intensive if high precision is sought. However, it presents little technical difficulty. Once the interpolating functions are constructed, further computations involving them are fast. We have shown that this approach is feasible and may be applied in practical spin hydrodynamics simulations.

We have also performed an analysis of the breakdown of the numerical simulations for large values of the spin polarization tensor in one of the two configurations considered (the second configuration being free from singularities). As discussed, singularities do form from smooth initial data also in other theories of relativistic hydrodynamics, such as the BDNK formalism, and the properties of global existence and global regularity of solutions do not automatically follow from stability and causality. Further works may show whether in the present framework the formation of singularities appears only on the perfect fluid level, in the theory expanded to the second order in $\omega$, and in the very restrictive geometry considered in this paper, or is it a more general limitation.

There are several possible interesting directions of further numerical study of spin hydrodynamics, such as: (I)
expanding the conserved current tensors up to a higher order in coefficients of $\omega$; (II) adding dissipation; (III)
considering spin-polarized fluids in
more general geometries, such as a boost-invariant cylindrically symmetric geometry and, ultimately, three-dimensional simulations analogous to those described in Refs.~\cite{Singh:2024cub,Sapna:2025yss}.
Each of those refinements may move spin hydrodynamics closer to a meaningful comparison with experimental data.

%%%%%%%%%%%%%%%%%%%%%%%%%%%%%%%%%%%%%%%%%%%%%%
%\bigskip
%\noindent
%{\it Data availability.} The data are available on request.

%%%%%%%%%%%%%%%%%%%%%%%%%%%%%%%%%%%%%%%%%%%%%%
\bigskip
\noindent

Acknowledgments -- The authors thank Wojciech Florkowski for the useful discussions and Radoslaw Ryblewski for sharing some Mathematica code. This work was supported by National Science Centre (NCN), Poland, Grant No. 2022/47/B/ST2/01372.

\bigskip

\begin{appendix}
\section{Conserved currents of second-order perfect spin hydrodynamics}\label{appendixA}

In relativistic hydrodynamics with a classical description of spin~\cite{Florkowski:2018fap,Drogosz:2024gzv},
the coefficients of the conserved current tensors are obtained by expanding to a given order (in the components of the spin polarization tensor $\omega_{\mu \nu}$) the~local equilibrium expressions 
\begin{align}\begin{split}
N^\mu_{\rm eq} &= \int \dd P \,\dd S \, p^\mu \, \left[f_{\rm eq}^+(x,p,s)-f_{\rm eq}^-(x,p,s) \right],\\
T^{\mu \nu}_{\rm eq} &= \int \dd P \,\dd S \, p^\mu p^\nu \, \left[f_{\rm eq}^+(x,p,s) + f_{\rm eq}^-(x,p,s) \right],\\
\hspace{-0.5cm}S^{\lambda \mu\nu}_{\rm eq} &= \int \!\dd P \, \dd S \, \, p^\lambda \, s^{\mu \nu} 
\left[f_{\rm eq}^+(x,p,s)+ f_{\rm eq}^-(x,p,s) \right],
\end{split}\end{align}
where the integration measures are 
\begin{equation}
\dd P = \frac{\dd^3 p}{(2\pi)^3 E_p}, \quad
\dd S = \frac{m}{\pi \s} \dd s \delta (s \cdot s + \s^2) \delta (p \cdot s),
\end{equation}
and $f_{\rm eq}^\pm(x,p,s)$ is either the Fermi–Dirac distribution for particles ($+$) or antiparticles ($-$)
\begin{equation}
f^{\pm}_{\rm eq}(x,p,s) = \frac{1}{\exp \Big( \mp \xi(x) + p \cdot \beta(x)  -  \frac{1}{2} \, \omega_{\mu \nu}(x) s^{\mu \nu} \Big)+1}
\end{equation}
or its Boltzmann approximation (particles only)
\begin{equation}
f_{\rm eq}(x,p,s) = \exp \big(\xi - p \cdot \beta(x) + \frac12 \omega_{\mu \nu}(x) s^{\mu \nu} \big).
\end{equation}
Expansions of the conserved currents up to the second order in the components of $\omega_{\mu \nu}$, derived this way in Ref.~\cite{Drogosz:2025ose}, are repeated for convenience in the notation of the present paper in Tables \ref{tab:S}–\ref{tab:N}. 
The tables use the following shorthand for sums or differences of the functions $\mathcal{J}_{mn}(\xi,z)$ of positive and negative first argument
\begin{align}\begin{split}
\mathcal{J}_{mnk+} \equiv \mathcal{J}_{mnk}(\xi,z), \quad \mathcal{J}_{mnk-} \equiv \mathcal{J}_{mnk}(-\xi,z),\\
\mathcal{J}_{mnk\Sigma} \equiv \mathcal{J}_{mnk+} + \mathcal{J}_{mnk-}, \quad \mathcal{J}_{mnk\Delta} \equiv \mathcal{J}_{mnk+} - \mathcal{J}_{mnk-}.
\end{split}\end{align}
The special functions $J_{mnk}(\xi,z)$ are defined by integrals \eqref{eq:js}--\eqref{eq:js2}, and their values have to be computed numerically. $K_j(z)$ is the modified Bessel function of the second kind.
For definitions of other symbols, see the main text.

{\everymath{\displaystyle}
\begin{table}[H]
\begin{center}
\caption{The spin tensor of perfect spin hydrodynamics with second-order spin corrections.}\label{tab:S}
\begin{tabular}{|ccc|}
\hline
\multicolumn{3}{|c|}{$\vph{10pt}{28pt}S_{\rm eq}^{\lambda \mu \nu} = u^\lambda\big[A(k^\mu u^\nu \!-\! k^\nu u^\mu) \!+\!A_1 t^{\mu\nu}\big] \!+\!\frac{A}{2}\big(t^{\lambda\mu} u^\nu \!-\! t^{\lambda \nu} u^\mu \!+ \!\Delta^{\lambda \mu} k^\nu \!-\! \Delta^{\lambda \nu}k^\mu \big)$} \\ \hline
\multicolumn{1}{|c|}{Coeff.} & \multicolumn{1}{c|}{Fermi--Dirac} & Boltzmann \\ \hline 
\multicolumn{1}{|c|}{$\vph{10pt}{28pt}A$} & \multicolumn{1}{c|}{$-\frac{2 \s^{2} m^3}{9 \pi^2} J_{411\Sigma}$} & $-\frac{4\s^{2}\!\cosh{\xi}}{3\pi^{2}}z T^{3}K_{3}(z)$ \\ \hline
\multicolumn{1}{|c|}{$\vph{10pt}{28pt}A_1$} & \multicolumn{1}{c|}{$\frac{\s^{2} m^3}{9 \pi^2} ( J_{211\Sigma} \!+\! 2 J_{231\Sigma})$} & $\frac{2\s^{2}\!\cosh{\xi}}{3\pi^{2}}zT^{3} \left[ zK_{2}(z)\!+\!2K_{3}(z)\right]$ \\ \hline
\end{tabular}
\end{center}
\end{table}
}

{\everymath{\displaystyle}
\begin{table}[H]
\begin{center}
\caption{The energy--momentum tensor of perfect spin hydrodynamics with second-order spin corrections.}\label{tab:T}
\begin{tabular}{|ccc|}
\hline
\multicolumn{3}{|c|}{$\begin{aligned}[t]
\vph{10pt}{24pt}T^{\mu\nu}_{\rm eq} &= (\varepsilon_0 + \varepsilon_2^\omega + \varepsilon_2^k)u^\mu u^\nu - (P_0 +P_2^\omega + P_2^k) \Delta^{\mu \nu}\\
&\vph{10pt}{18pt}+ (t^\mu u^\nu + t^\nu u^\mu)P_t + (k^\mu k^\nu + \omega^\mu \omega ^\nu)P_{k\omega}  \end{aligned}$}                            \\ \hline
\multicolumn{1}{|c|}{Coeff.} & \multicolumn{1}{c|}{Fermi--Dirac} & Boltzmann \\ \hline
\multicolumn{1}{|c|}{$\vph{10pt}{26pt}\varepsilon_0$} & \multicolumn{1}{c|}{$\frac{m^4}{\pi^2 } J_{220\Sigma} $} & $\frac{2 \cosh \xi}{\pi^2} z^2 T^4 [z K_3(z) \!-\! K_2(z)]$ \\ \hline
\multicolumn{1}{|c|}{$\vph{10pt}{28pt}\varepsilon_2^\omega$} & \multicolumn{1}{c|}{$- \omega^2 \frac{\s^2 m^4}{18\pi^2}(J_{222\Sigma} \!+\! 2 J_{242\Sigma})$} & $\begin{aligned}[t]
&- \omega^2\frac{\s^2 \!\cosh \xi}{3 \pi^2} z T^4 \, \big[zK_2(z) \\
&\quad+ (z^2 \!+\! 10) K_3(z) \big] \end{aligned}$ \\ \hline
\multicolumn{1}{|c|}{$\vph{10pt}{28pt}\varepsilon_2^k$} & \multicolumn{1}{c|}{$- k^2 \frac{\s^2 m^4}{9\pi^2} J_{422\Sigma}$} & \multicolumn{1}{@{\hspace{3pt}}c@{\hspace{3pt}}|}{$-k^2 \frac{2 \s^2 \!\cosh \xi}{3\pi^2}zT^4[zK_2(z) \!+\! 5K_3(z)]$} \\ \hline
\multicolumn{1}{|c|}{$\vph{10pt}{26pt}P_0$} & \multicolumn{1}{c|}{$\frac{m^4}{3\pi^2} J_{400\Sigma}$} & $\frac{2 \cosh \xi}{\pi^2}z^2 T^4 K_2(z)$ \\ \hline
\multicolumn{1}{|c|}{$\vph{10pt}{28pt}P_2^\omega$} & \multicolumn{1}{c|}{$- \omega^2 \frac{\s^2 m^4}{90 \pi^2} (J_{402\Sigma} \!+\! 4 J_{422\Sigma})$} & \multicolumn{1}{@{\hspace{3pt}}c@{\hspace{3pt}}|}{$-\omega^2 \frac{\s^2 \!\cosh \xi}{3 \pi^2} zT^4 [zK_2(z) \!+\! 4K_3(z)] $} \\ \hline
\multicolumn{1}{|c|}{$\vph{10pt}{28pt}P_2^k$} & \multicolumn{1}{c|}{$- k^2 \frac{2\s^2 m^4}{45 \pi^2} J_{602\Sigma}$} & $- k^2 \frac{4 \s^2 \!\cosh \xi}{3 \pi^2}zT^4 K_3(z)$ \\ \hline
\multicolumn{1}{|c|}{$\vph{10pt}{28pt}P_{k \omega}$} & \multicolumn{1}{c|}{$-\frac{\s^2 m^4}{45 \pi^2} J_{602\Sigma} $} & $-\frac{2\s^2 \!\cosh \xi}{3 \pi^2}zT^4 K_3(z)$ \\ \hline
\multicolumn{1}{|c|}{$\vph{10pt}{28pt}P_{t}$} & \multicolumn{1}{c|}{$-\frac{\s^2m^4}{9\pi^2}J_{422\Sigma}$} & $-\frac{2 \s^2 \!\cosh \xi}{3 \pi^2} zT^4 [zK_2(z) \!+\! 5K_3(z)]$ \\ \hline
\end{tabular}
\end{center}
\end{table}
}

{\everymath{\displaystyle}
\begin{table}[H]
\begin{center}
\caption{The baryon current of perfect spin hydrodynamics with second-order spin corrections.}\label{tab:N}
\begin{tabular}{|ccc|}
\hline
\multicolumn{3}{|c|}{$\vphantom{\bigg|}N_{\rm eq}^\mu = ( n_0+ n_2^\omega +  n_2^k )u^{\mu}+n_{t}t^{\mu}$}                            \\ \hline
\multicolumn{1}{|c|}{Coeff.} & \multicolumn{1}{c|}{Fermi--Dirac} & Boltzmann \\ \hline 
\multicolumn{1}{|c|}{$\vph{10pt}{26pt}n_0$} & \multicolumn{1}{c|}{$\frac{m^3}{\pi^2} J_{210\Delta}$} & $\frac{2\sinh{\xi}}{\pi^{2}}z^{2}T^{3}K_{2}(z)$ \\ \hline
\multicolumn{1}{|c|}{$\vph{10pt}{28pt}n_2^\omega$} & \multicolumn{1}{c|}{$ - \omega^2 \frac{\s^2 m^3}{18 \pi^2} (J_{212\Delta} \!+\! 2 J_{232\Delta} )$} & $- \omega^2 \frac{\s^{2}\!\sinh{\xi}}{3\pi^{2}} z T^{3} \left[ z K_{2}(z) \!+\! 2 K_{3}(z) \right]$ \\ \hline
\multicolumn{1}{|c|}{$\vph{10pt}{28pt}n_2^k$} & \multicolumn{1}{c|}{$- k^2 \frac{\s^2 m^3}{9\pi^2} J_{412\Delta}$} & $-k^2\frac{2 \s^{2} \!\sinh{\xi}}{3\pi^{2}} z T^{3}K_{3}(z)$ \\ \hline
\multicolumn{1}{|c|}{$\vph{10pt}{28pt}n_t$} & \multicolumn{1}{c|}{$- \frac{\s^2 m^3}{9\pi^2} J_{412\Delta}$} & $-\frac{2\s^{2}\!\sinh{\xi}}{3\pi^2}  z T^{3} K_{3}(z)$ \\ \hline
\end{tabular}
\end{center}
\end{table}
}

\end{appendix}

%\printbibliography
%\bibliography{fd-boost-inv}

%\bibliographystyle{apsrev4-1}
%\bibliography{fd-boost-inv.bib}

\begin{thebibliography}{61}%
\makeatletter
\providecommand \@ifxundefined [1]{%
 \@ifx{#1\undefined}
}%
\providecommand \@ifnum [1]{%
 \ifnum #1\expandafter \@firstoftwo
 \else \expandafter \@secondoftwo
 \fi
}%
\providecommand \@ifx [1]{%
 \ifx #1\expandafter \@firstoftwo
 \else \expandafter \@secondoftwo
 \fi
}%
\providecommand \natexlab [1]{#1}%
\providecommand \enquote  [1]{``#1''}%
\providecommand \bibnamefont  [1]{#1}%
\providecommand \bibfnamefont [1]{#1}%
\providecommand \citenamefont [1]{#1}%
\providecommand \href@noop [0]{\@secondoftwo}%
\providecommand \href [0]{\begingroup \@sanitize@url \@href}%
\providecommand \@href[1]{\@@startlink{#1}\@@href}%
\providecommand \@@href[1]{\endgroup#1\@@endlink}%
\providecommand \@sanitize@url [0]{\catcode `\\12\catcode `\$12\catcode
  `\&12\catcode `\#12\catcode `\^12\catcode `\_12\catcode `\%12\relax}%
\providecommand \@@startlink[1]{}%
\providecommand \@@endlink[0]{}%
\providecommand \url  [0]{\begingroup\@sanitize@url \@url }%
\providecommand \@url [1]{\endgroup\@href {#1}{\urlprefix }}%
\providecommand \urlprefix  [0]{URL }%
\providecommand \Eprint [0]{\href }%
\providecommand \doibase [0]{http://dx.doi.org/}%
\providecommand \selectlanguage [0]{\@gobble}%
\providecommand \bibinfo  [0]{\@secondoftwo}%
\providecommand \bibfield  [0]{\@secondoftwo}%
\providecommand \translation [1]{[#1]}%
\providecommand \BibitemOpen [0]{}%
\providecommand \bibitemStop [0]{}%
\providecommand \bibitemNoStop [0]{.\EOS\space}%
\providecommand \EOS [0]{\spacefactor3000\relax}%
\providecommand \BibitemShut  [1]{\csname bibitem#1\endcsname}%
\let\auto@bib@innerbib\@empty
%</preamble>
\bibitem [{\citenamefont {Niida}\ and\ \citenamefont
  {Voloshin}(2024)}]{Niida:2024ntm}%
  \BibitemOpen
  \bibfield  {author} {\bibinfo {author} {\bibfnamefont {T.}~\bibnamefont
  {Niida}}\ and\ \bibinfo {author} {\bibfnamefont {S.~A.}\ \bibnamefont
  {Voloshin}},\ }\href {\doibase 10.1142/S0218301324300108} {\bibfield
  {journal} {\bibinfo  {journal} {Int. J. Mod. Phys. E}\ }\textbf {\bibinfo
  {volume} {33}},\ \bibinfo {pages} {2430010} (\bibinfo {year} {2024})},\
  \Eprint {http://arxiv.org/abs/2404.11042} {arXiv:2404.11042 [nucl-ex]}
  \BibitemShut {NoStop}%
\bibitem [{\citenamefont {Adamczyk}\ \emph {et~al.}(2017)\citenamefont
  {Adamczyk} \emph {et~al.}}]{STAR:2017ckg}%
  \BibitemOpen
  \bibfield  {author} {\bibinfo {author} {\bibfnamefont {L.}~\bibnamefont
  {Adamczyk}} \emph {et~al.} (\bibinfo {collaboration} {STAR}),\ }\href
  {\doibase 10.1038/nature23004} {\bibfield  {journal} {\bibinfo  {journal}
  {Nature}\ }\textbf {\bibinfo {volume} {548}},\ \bibinfo {pages} {62}
  (\bibinfo {year} {2017})},\ \Eprint {http://arxiv.org/abs/1701.06657}
  {arXiv:1701.06657 [nucl-ex]} \BibitemShut {NoStop}%
\bibitem [{\citenamefont {Adam}\ \emph {et~al.}(2018)\citenamefont {Adam} \emph
  {et~al.}}]{STAR:2018gyt}%
  \BibitemOpen
  \bibfield  {author} {\bibinfo {author} {\bibfnamefont {J.}~\bibnamefont
  {Adam}} \emph {et~al.} (\bibinfo {collaboration} {STAR}),\ }\href {\doibase
  10.1103/PhysRevC.98.014910} {\bibfield  {journal} {\bibinfo  {journal} {Phys.
  Rev. C}\ }\textbf {\bibinfo {volume} {98}},\ \bibinfo {pages} {014910}
  (\bibinfo {year} {2018})},\ \Eprint {http://arxiv.org/abs/1805.04400}
  {arXiv:1805.04400 [nucl-ex]} \BibitemShut {NoStop}%
\bibitem [{\citenamefont {Adam}\ \emph {et~al.}(2019)\citenamefont {Adam} \emph
  {et~al.}}]{STAR:2019erd}%
  \BibitemOpen
  \bibfield  {author} {\bibinfo {author} {\bibfnamefont {J.}~\bibnamefont
  {Adam}} \emph {et~al.} (\bibinfo {collaboration} {STAR}),\ }\href {\doibase
  10.1103/PhysRevLett.123.132301} {\bibfield  {journal} {\bibinfo  {journal}
  {Phys. Rev. Lett.}\ }\textbf {\bibinfo {volume} {123}},\ \bibinfo {pages}
  {132301} (\bibinfo {year} {2019})},\ \Eprint
  {http://arxiv.org/abs/1905.11917} {arXiv:1905.11917 [nucl-ex]} \BibitemShut
  {NoStop}%
\bibitem [{\citenamefont {Acharya}\ \emph {et~al.}(2020)\citenamefont {Acharya}
  \emph {et~al.}}]{ALICE:2019aid}%
  \BibitemOpen
  \bibfield  {author} {\bibinfo {author} {\bibfnamefont {S.}~\bibnamefont
  {Acharya}} \emph {et~al.} (\bibinfo {collaboration} {ALICE}),\ }\href
  {\doibase 10.1103/PhysRevLett.125.012301} {\bibfield  {journal} {\bibinfo
  {journal} {Phys. Rev. Lett.}\ }\textbf {\bibinfo {volume} {125}},\ \bibinfo
  {pages} {012301} (\bibinfo {year} {2020})},\ \Eprint
  {http://arxiv.org/abs/1910.14408} {arXiv:1910.14408 [nucl-ex]} \BibitemShut
  {NoStop}%
\bibitem [{\citenamefont {Becattini}\ and\ \citenamefont
  {Tinti}(2010)}]{Becattini:2009wh}%
  \BibitemOpen
  \bibfield  {author} {\bibinfo {author} {\bibfnamefont {F.}~\bibnamefont
  {Becattini}}\ and\ \bibinfo {author} {\bibfnamefont {L.}~\bibnamefont
  {Tinti}},\ }\href {\doibase 10.1016/j.aop.2010.03.007} {\bibfield  {journal}
  {\bibinfo  {journal} {Annals Phys.}\ }\textbf {\bibinfo {volume} {325}},\
  \bibinfo {pages} {1566} (\bibinfo {year} {2010})},\ \Eprint
  {http://arxiv.org/abs/0911.0864} {arXiv:0911.0864 [gr-qc]} \BibitemShut
  {NoStop}%
\bibitem [{\citenamefont {Becattini}\ \emph {et~al.}(2021)\citenamefont
  {Becattini}, \citenamefont {Buzzegoli}, \citenamefont {Inghirami},
  \citenamefont {Karpenko},\ and\ \citenamefont {Palermo}}]{Becattini:2021iol}%
  \BibitemOpen
  \bibfield  {author} {\bibinfo {author} {\bibfnamefont {F.}~\bibnamefont
  {Becattini}}, \bibinfo {author} {\bibfnamefont {M.}~\bibnamefont
  {Buzzegoli}}, \bibinfo {author} {\bibfnamefont {G.}~\bibnamefont
  {Inghirami}}, \bibinfo {author} {\bibfnamefont {I.}~\bibnamefont {Karpenko}},
  \ and\ \bibinfo {author} {\bibfnamefont {A.}~\bibnamefont {Palermo}},\ }\href
  {\doibase 10.1103/PhysRevLett.127.272302} {\bibfield  {journal} {\bibinfo
  {journal} {Phys. Rev. Lett.}\ }\textbf {\bibinfo {volume} {127}},\ \bibinfo
  {pages} {272302} (\bibinfo {year} {2021})},\ \Eprint
  {http://arxiv.org/abs/2103.14621} {arXiv:2103.14621 [nucl-th]} \BibitemShut
  {NoStop}%
\bibitem [{\citenamefont {Palermo}\ \emph {et~al.}(2024)\citenamefont
  {Palermo}, \citenamefont {Grossi}, \citenamefont {Karpenko},\ and\
  \citenamefont {Becattini}}]{Palermo:2024tza}%
  \BibitemOpen
  \bibfield  {author} {\bibinfo {author} {\bibfnamefont {A.}~\bibnamefont
  {Palermo}}, \bibinfo {author} {\bibfnamefont {E.}~\bibnamefont {Grossi}},
  \bibinfo {author} {\bibfnamefont {I.}~\bibnamefont {Karpenko}}, \ and\
  \bibinfo {author} {\bibfnamefont {F.}~\bibnamefont {Becattini}},\ }\href
  {\doibase 10.1140/epjc/s10052-024-13229-z} {\bibfield  {journal} {\bibinfo
  {journal} {Eur. Phys. J. C}\ }\textbf {\bibinfo {volume} {84}},\ \bibinfo
  {pages} {920} (\bibinfo {year} {2024})},\ \Eprint
  {http://arxiv.org/abs/2404.14295} {arXiv:2404.14295 [nucl-th]} \BibitemShut
  {NoStop}%
\bibitem [{\citenamefont {Florkowski}\ \emph
  {et~al.}(2018{\natexlab{a}})\citenamefont {Florkowski}, \citenamefont
  {Friman}, \citenamefont {Jaiswal},\ and\ \citenamefont
  {Speranza}}]{Florkowski:2017ruc}%
  \BibitemOpen
  \bibfield  {author} {\bibinfo {author} {\bibfnamefont {W.}~\bibnamefont
  {Florkowski}}, \bibinfo {author} {\bibfnamefont {B.}~\bibnamefont {Friman}},
  \bibinfo {author} {\bibfnamefont {A.}~\bibnamefont {Jaiswal}}, \ and\
  \bibinfo {author} {\bibfnamefont {E.}~\bibnamefont {Speranza}},\ }\href
  {\doibase 10.1103/PhysRevC.97.041901} {\bibfield  {journal} {\bibinfo
  {journal} {Phys. Rev.}\ }\textbf {\bibinfo {volume} {C97}},\ \bibinfo {pages}
  {041901} (\bibinfo {year} {2018}{\natexlab{a}})},\ \Eprint
  {http://arxiv.org/abs/1705.00587} {arXiv:1705.00587 [nucl-th]} \BibitemShut
  {NoStop}%
%%CITATION = ARXIV:1705.00587;%%
\bibitem [{\citenamefont {Shi}\ \emph {et~al.}(2021)\citenamefont {Shi},
  \citenamefont {Gale},\ and\ \citenamefont {Jeon}}]{Shi:2020htn}%
  \BibitemOpen
  \bibfield  {author} {\bibinfo {author} {\bibfnamefont {S.}~\bibnamefont
  {Shi}}, \bibinfo {author} {\bibfnamefont {C.}~\bibnamefont {Gale}}, \ and\
  \bibinfo {author} {\bibfnamefont {S.}~\bibnamefont {Jeon}},\ }\href {\doibase
  10.1103/PhysRevC.103.044906} {\bibfield  {journal} {\bibinfo  {journal}
  {Phys. Rev. C}\ }\textbf {\bibinfo {volume} {103}},\ \bibinfo {pages}
  {044906} (\bibinfo {year} {2021})},\ \Eprint
  {http://arxiv.org/abs/2008.08618} {arXiv:2008.08618 [nucl-th]} \BibitemShut
  {NoStop}%
\bibitem [{\citenamefont {Hu}(2022)}]{Hu:2021pwh}%
  \BibitemOpen
  \bibfield  {author} {\bibinfo {author} {\bibfnamefont {J.}~\bibnamefont
  {Hu}},\ }\href {\doibase 10.1103/PhysRevD.105.076009} {\bibfield  {journal}
  {\bibinfo  {journal} {Phys. Rev. D}\ }\textbf {\bibinfo {volume} {105}},\
  \bibinfo {pages} {076009} (\bibinfo {year} {2022})},\ \Eprint
  {http://arxiv.org/abs/2111.03571} {arXiv:2111.03571 [hep-ph]} \BibitemShut
  {NoStop}%
\bibitem [{\citenamefont {Bhadury}\ \emph {et~al.}(2021)\citenamefont
  {Bhadury}, \citenamefont {Florkowski}, \citenamefont {Jaiswal}, \citenamefont
  {Kumar},\ and\ \citenamefont {Ryblewski}}]{Bhadury:2020puc}%
  \BibitemOpen
  \bibfield  {author} {\bibinfo {author} {\bibfnamefont {S.}~\bibnamefont
  {Bhadury}}, \bibinfo {author} {\bibfnamefont {W.}~\bibnamefont {Florkowski}},
  \bibinfo {author} {\bibfnamefont {A.}~\bibnamefont {Jaiswal}}, \bibinfo
  {author} {\bibfnamefont {A.}~\bibnamefont {Kumar}}, \ and\ \bibinfo {author}
  {\bibfnamefont {R.}~\bibnamefont {Ryblewski}},\ }\href {\doibase
  10.1016/j.physletb.2021.136096} {\bibfield  {journal} {\bibinfo  {journal}
  {Phys. Lett. B}\ }\textbf {\bibinfo {volume} {814}},\ \bibinfo {pages}
  {136096} (\bibinfo {year} {2021})},\ \Eprint
  {http://arxiv.org/abs/2002.03937} {arXiv:2002.03937 [hep-ph]} \BibitemShut
  {NoStop}%
\bibitem [{\citenamefont {Bhadury}\ \emph {et~al.}(2022)\citenamefont
  {Bhadury}, \citenamefont {Florkowski}, \citenamefont {Jaiswal}, \citenamefont
  {Kumar},\ and\ \citenamefont {Ryblewski}}]{Bhadury:2022ulr}%
  \BibitemOpen
  \bibfield  {author} {\bibinfo {author} {\bibfnamefont {S.}~\bibnamefont
  {Bhadury}}, \bibinfo {author} {\bibfnamefont {W.}~\bibnamefont {Florkowski}},
  \bibinfo {author} {\bibfnamefont {A.}~\bibnamefont {Jaiswal}}, \bibinfo
  {author} {\bibfnamefont {A.}~\bibnamefont {Kumar}}, \ and\ \bibinfo {author}
  {\bibfnamefont {R.}~\bibnamefont {Ryblewski}},\ }\href {\doibase
  10.1103/PhysRevLett.129.192301} {\bibfield  {journal} {\bibinfo  {journal}
  {Phys. Rev. Lett.}\ }\textbf {\bibinfo {volume} {129}},\ \bibinfo {pages}
  {192301} (\bibinfo {year} {2022})},\ \Eprint
  {http://arxiv.org/abs/2204.01357} {arXiv:2204.01357 [nucl-th]} \BibitemShut
  {NoStop}%
\bibitem [{\citenamefont {Weickgenannt}\ \emph {et~al.}(2019)\citenamefont
  {Weickgenannt}, \citenamefont {Sheng}, \citenamefont {Speranza},
  \citenamefont {Wang},\ and\ \citenamefont {Rischke}}]{Weickgenannt:2019dks}%
  \BibitemOpen
  \bibfield  {author} {\bibinfo {author} {\bibfnamefont {N.}~\bibnamefont
  {Weickgenannt}}, \bibinfo {author} {\bibfnamefont {X.-L.}\ \bibnamefont
  {Sheng}}, \bibinfo {author} {\bibfnamefont {E.}~\bibnamefont {Speranza}},
  \bibinfo {author} {\bibfnamefont {Q.}~\bibnamefont {Wang}}, \ and\ \bibinfo
  {author} {\bibfnamefont {D.~H.}\ \bibnamefont {Rischke}},\ }\href {\doibase
  10.1103/PhysRevD.100.056018} {\bibfield  {journal} {\bibinfo  {journal}
  {Phys. Rev. D}\ }\textbf {\bibinfo {volume} {100}},\ \bibinfo {pages}
  {056018} (\bibinfo {year} {2019})},\ \Eprint
  {http://arxiv.org/abs/1902.06513} {arXiv:1902.06513 [hep-ph]} \BibitemShut
  {NoStop}%
\bibitem [{\citenamefont {Weickgenannt}\ \emph
  {et~al.}(2021{\natexlab{a}})\citenamefont {Weickgenannt}, \citenamefont
  {Speranza}, \citenamefont {Sheng}, \citenamefont {Wang},\ and\ \citenamefont
  {Rischke}}]{Weickgenannt:2020aaf}%
  \BibitemOpen
  \bibfield  {author} {\bibinfo {author} {\bibfnamefont {N.}~\bibnamefont
  {Weickgenannt}}, \bibinfo {author} {\bibfnamefont {E.}~\bibnamefont
  {Speranza}}, \bibinfo {author} {\bibfnamefont {X.-l.}\ \bibnamefont {Sheng}},
  \bibinfo {author} {\bibfnamefont {Q.}~\bibnamefont {Wang}}, \ and\ \bibinfo
  {author} {\bibfnamefont {D.~H.}\ \bibnamefont {Rischke}},\ }\href {\doibase
  10.1103/PhysRevLett.127.052301} {\bibfield  {journal} {\bibinfo  {journal}
  {Phys. Rev. Lett.}\ }\textbf {\bibinfo {volume} {127}},\ \bibinfo {pages}
  {052301} (\bibinfo {year} {2021}{\natexlab{a}})},\ \Eprint
  {http://arxiv.org/abs/2005.01506} {arXiv:2005.01506 [hep-ph]} \BibitemShut
  {NoStop}%
\bibitem [{\citenamefont {Weickgenannt}\ \emph
  {et~al.}(2021{\natexlab{b}})\citenamefont {Weickgenannt}, \citenamefont
  {Speranza}, \citenamefont {Sheng}, \citenamefont {Wang},\ and\ \citenamefont
  {Rischke}}]{Weickgenannt:2021cuo}%
  \BibitemOpen
  \bibfield  {author} {\bibinfo {author} {\bibfnamefont {N.}~\bibnamefont
  {Weickgenannt}}, \bibinfo {author} {\bibfnamefont {E.}~\bibnamefont
  {Speranza}}, \bibinfo {author} {\bibfnamefont {X.-l.}\ \bibnamefont {Sheng}},
  \bibinfo {author} {\bibfnamefont {Q.}~\bibnamefont {Wang}}, \ and\ \bibinfo
  {author} {\bibfnamefont {D.~H.}\ \bibnamefont {Rischke}},\ }\href {\doibase
  10.1103/PhysRevD.104.016022} {\bibfield  {journal} {\bibinfo  {journal}
  {Phys. Rev. D}\ }\textbf {\bibinfo {volume} {104}},\ \bibinfo {pages}
  {016022} (\bibinfo {year} {2021}{\natexlab{b}})},\ \Eprint
  {http://arxiv.org/abs/2103.04896} {arXiv:2103.04896 [nucl-th]} \BibitemShut
  {NoStop}%
\bibitem [{\citenamefont {Weickgenannt}\ \emph {et~al.}(2022)\citenamefont
  {Weickgenannt}, \citenamefont {Wagner}, \citenamefont {Speranza},\ and\
  \citenamefont {Rischke}}]{Weickgenannt:2022zxs}%
  \BibitemOpen
  \bibfield  {author} {\bibinfo {author} {\bibfnamefont {N.}~\bibnamefont
  {Weickgenannt}}, \bibinfo {author} {\bibfnamefont {D.}~\bibnamefont
  {Wagner}}, \bibinfo {author} {\bibfnamefont {E.}~\bibnamefont {Speranza}}, \
  and\ \bibinfo {author} {\bibfnamefont {D.~H.}\ \bibnamefont {Rischke}},\
  }\href {\doibase 10.1103/PhysRevD.106.096014} {\bibfield  {journal} {\bibinfo
   {journal} {Phys. Rev. D}\ }\textbf {\bibinfo {volume} {106}},\ \bibinfo
  {pages} {096014} (\bibinfo {year} {2022})},\ \Eprint
  {http://arxiv.org/abs/2203.04766} {arXiv:2203.04766 [nucl-th]} \BibitemShut
  {NoStop}%
\bibitem [{\citenamefont {Weickgenannt}\ and\ \citenamefont
  {Blaizot}(2024)}]{Weickgenannt:2023nge}%
  \BibitemOpen
  \bibfield  {author} {\bibinfo {author} {\bibfnamefont {N.}~\bibnamefont
  {Weickgenannt}}\ and\ \bibinfo {author} {\bibfnamefont {J.-P.}\ \bibnamefont
  {Blaizot}},\ }\href {\doibase 10.1103/PhysRevD.109.056012} {\bibfield
  {journal} {\bibinfo  {journal} {Phys. Rev. D}\ }\textbf {\bibinfo {volume}
  {109}},\ \bibinfo {pages} {056012} (\bibinfo {year} {2024})},\ \Eprint
  {http://arxiv.org/abs/2311.15817} {arXiv:2311.15817 [hep-ph]} \BibitemShut
  {NoStop}%
\bibitem [{\citenamefont {Wagner}\ \emph {et~al.}(2024)\citenamefont {Wagner},
  \citenamefont {Shokri},\ and\ \citenamefont {Rischke}}]{Wagner:2024fhf}%
  \BibitemOpen
  \bibfield  {author} {\bibinfo {author} {\bibfnamefont {D.}~\bibnamefont
  {Wagner}}, \bibinfo {author} {\bibfnamefont {M.}~\bibnamefont {Shokri}}, \
  and\ \bibinfo {author} {\bibfnamefont {D.~H.}\ \bibnamefont {Rischke}},\
  }\href {\doibase 10.1103/PhysRevResearch.6.043103} {\bibfield  {journal}
  {\bibinfo  {journal} {Phys. Rev. Res.}\ }\textbf {\bibinfo {volume} {6}},\
  \bibinfo {pages} {043103} (\bibinfo {year} {2024})},\ \Eprint
  {http://arxiv.org/abs/2405.00533} {arXiv:2405.00533 [nucl-th]} \BibitemShut
  {NoStop}%
\bibitem [{\citenamefont {Banerjee}\ \emph {et~al.}(2025)\citenamefont
  {Banerjee}, \citenamefont {Bhadury}, \citenamefont {Florkowski},
  \citenamefont {Jaiswal},\ and\ \citenamefont {Ryblewski}}]{Banerjee:2024xnd}%
  \BibitemOpen
  \bibfield  {author} {\bibinfo {author} {\bibfnamefont {S.}~\bibnamefont
  {Banerjee}}, \bibinfo {author} {\bibfnamefont {S.}~\bibnamefont {Bhadury}},
  \bibinfo {author} {\bibfnamefont {W.}~\bibnamefont {Florkowski}}, \bibinfo
  {author} {\bibfnamefont {A.}~\bibnamefont {Jaiswal}}, \ and\ \bibinfo
  {author} {\bibfnamefont {R.}~\bibnamefont {Ryblewski}},\ }\href {\doibase
  10.1103/923l-yxkc} {\bibfield  {journal} {\bibinfo  {journal} {Phys. Rev. C}\
  }\textbf {\bibinfo {volume} {111}},\ \bibinfo {pages} {064912} (\bibinfo
  {year} {2025})},\ \Eprint {http://arxiv.org/abs/2405.05089} {arXiv:2405.05089
  [hep-ph]} \BibitemShut {NoStop}%
\bibitem [{\citenamefont {Bhadury}(2025)}]{Bhadury:2024ckc}%
  \BibitemOpen
  \bibfield  {author} {\bibinfo {author} {\bibfnamefont {S.}~\bibnamefont
  {Bhadury}},\ }\href {\doibase 10.1103/PhysRevC.111.034909} {\bibfield
  {journal} {\bibinfo  {journal} {Phys. Rev. C}\ }\textbf {\bibinfo {volume}
  {111}},\ \bibinfo {pages} {034909} (\bibinfo {year} {2025})},\ \Eprint
  {http://arxiv.org/abs/2408.14462} {arXiv:2408.14462 [hep-ph]} \BibitemShut
  {NoStop}%
\bibitem [{\citenamefont {Li}\ \emph {et~al.}(2021)\citenamefont {Li},
  \citenamefont {Stephanov},\ and\ \citenamefont {Yee}}]{Li:2020eon}%
  \BibitemOpen
  \bibfield  {author} {\bibinfo {author} {\bibfnamefont {S.}~\bibnamefont
  {Li}}, \bibinfo {author} {\bibfnamefont {M.~A.}\ \bibnamefont {Stephanov}}, \
  and\ \bibinfo {author} {\bibfnamefont {H.-U.}\ \bibnamefont {Yee}},\ }\href
  {\doibase 10.1103/PhysRevLett.127.082302} {\bibfield  {journal} {\bibinfo
  {journal} {Phys. Rev. Lett.}\ }\textbf {\bibinfo {volume} {127}},\ \bibinfo
  {pages} {082302} (\bibinfo {year} {2021})},\ \Eprint
  {http://arxiv.org/abs/2011.12318} {arXiv:2011.12318 [hep-th]} \BibitemShut
  {NoStop}%
\bibitem [{\citenamefont {Hattori}\ \emph {et~al.}(2019)\citenamefont
  {Hattori}, \citenamefont {Hongo}, \citenamefont {Huang}, \citenamefont
  {Matsuo},\ and\ \citenamefont {Taya}}]{Hattori:2019lfp}%
  \BibitemOpen
  \bibfield  {author} {\bibinfo {author} {\bibfnamefont {K.}~\bibnamefont
  {Hattori}}, \bibinfo {author} {\bibfnamefont {M.}~\bibnamefont {Hongo}},
  \bibinfo {author} {\bibfnamefont {X.-G.}\ \bibnamefont {Huang}}, \bibinfo
  {author} {\bibfnamefont {M.}~\bibnamefont {Matsuo}}, \ and\ \bibinfo {author}
  {\bibfnamefont {H.}~\bibnamefont {Taya}},\ }\href {\doibase
  10.1016/j.physletb.2019.05.040} {\bibfield  {journal} {\bibinfo  {journal}
  {Phys. Lett. B}\ }\textbf {\bibinfo {volume} {795}},\ \bibinfo {pages} {100}
  (\bibinfo {year} {2019})},\ \Eprint {http://arxiv.org/abs/1901.06615}
  {arXiv:1901.06615 [hep-th]} \BibitemShut {NoStop}%
\bibitem [{\citenamefont {Fukushima}\ and\ \citenamefont
  {Pu}(2021)}]{Fukushima:2020ucl}%
  \BibitemOpen
  \bibfield  {author} {\bibinfo {author} {\bibfnamefont {K.}~\bibnamefont
  {Fukushima}}\ and\ \bibinfo {author} {\bibfnamefont {S.}~\bibnamefont {Pu}},\
  }\href {\doibase 10.1016/j.physletb.2021.136346} {\bibfield  {journal}
  {\bibinfo  {journal} {Phys. Lett. B}\ }\textbf {\bibinfo {volume} {817}},\
  \bibinfo {pages} {136346} (\bibinfo {year} {2021})},\ \Eprint
  {http://arxiv.org/abs/2010.01608} {arXiv:2010.01608 [hep-th]} \BibitemShut
  {NoStop}%
\bibitem [{\citenamefont {Biswas}\ \emph {et~al.}(2023)\citenamefont {Biswas},
  \citenamefont {Daher}, \citenamefont {Das}, \citenamefont {Florkowski},\ and\
  \citenamefont {Ryblewski}}]{Biswas:2023qsw}%
  \BibitemOpen
  \bibfield  {author} {\bibinfo {author} {\bibfnamefont {R.}~\bibnamefont
  {Biswas}}, \bibinfo {author} {\bibfnamefont {A.}~\bibnamefont {Daher}},
  \bibinfo {author} {\bibfnamefont {A.}~\bibnamefont {Das}}, \bibinfo {author}
  {\bibfnamefont {W.}~\bibnamefont {Florkowski}}, \ and\ \bibinfo {author}
  {\bibfnamefont {R.}~\bibnamefont {Ryblewski}},\ }\href {\doibase
  10.1103/PhysRevD.108.014024} {\bibfield  {journal} {\bibinfo  {journal}
  {Phys. Rev. D}\ }\textbf {\bibinfo {volume} {108}},\ \bibinfo {pages}
  {014024} (\bibinfo {year} {2023})},\ \Eprint
  {http://arxiv.org/abs/2304.01009} {arXiv:2304.01009 [nucl-th]} \BibitemShut
  {NoStop}%
\bibitem [{\citenamefont {Xie}\ \emph {et~al.}(2023)\citenamefont {Xie},
  \citenamefont {Wang}, \citenamefont {Yang},\ and\ \citenamefont
  {Pu}}]{Xie:2023gbo}%
  \BibitemOpen
  \bibfield  {author} {\bibinfo {author} {\bibfnamefont {X.-Q.}\ \bibnamefont
  {Xie}}, \bibinfo {author} {\bibfnamefont {D.-L.}\ \bibnamefont {Wang}},
  \bibinfo {author} {\bibfnamefont {C.}~\bibnamefont {Yang}}, \ and\ \bibinfo
  {author} {\bibfnamefont {S.}~\bibnamefont {Pu}},\ }\href {\doibase
  10.1103/PhysRevD.108.094031} {\bibfield  {journal} {\bibinfo  {journal}
  {Phys. Rev. D}\ }\textbf {\bibinfo {volume} {108}},\ \bibinfo {pages}
  {094031} (\bibinfo {year} {2023})},\ \Eprint
  {http://arxiv.org/abs/2306.13880} {arXiv:2306.13880 [hep-ph]} \BibitemShut
  {NoStop}%
\bibitem [{\citenamefont {Daher}\ \emph
  {et~al.}(2024{\natexlab{a}})\citenamefont {Daher}, \citenamefont
  {Florkowski},\ and\ \citenamefont {Ryblewski}}]{Daher:2024ixz}%
  \BibitemOpen
  \bibfield  {author} {\bibinfo {author} {\bibfnamefont {A.}~\bibnamefont
  {Daher}}, \bibinfo {author} {\bibfnamefont {W.}~\bibnamefont {Florkowski}}, \
  and\ \bibinfo {author} {\bibfnamefont {R.}~\bibnamefont {Ryblewski}},\ }\href
  {\doibase 10.1103/PhysRevD.110.034029} {\bibfield  {journal} {\bibinfo
  {journal} {Phys. Rev. D}\ }\textbf {\bibinfo {volume} {110}},\ \bibinfo
  {pages} {034029} (\bibinfo {year} {2024}{\natexlab{a}})},\ \Eprint
  {http://arxiv.org/abs/2401.07608} {arXiv:2401.07608 [hep-ph]} \BibitemShut
  {NoStop}%
\bibitem [{\citenamefont {Ren}\ \emph {et~al.}(2024)\citenamefont {Ren},
  \citenamefont {Yang}, \citenamefont {Wang},\ and\ \citenamefont
  {Pu}}]{Ren:2024pur}%
  \BibitemOpen
  \bibfield  {author} {\bibinfo {author} {\bibfnamefont {X.}~\bibnamefont
  {Ren}}, \bibinfo {author} {\bibfnamefont {C.}~\bibnamefont {Yang}}, \bibinfo
  {author} {\bibfnamefont {D.-L.}\ \bibnamefont {Wang}}, \ and\ \bibinfo
  {author} {\bibfnamefont {S.}~\bibnamefont {Pu}},\ }\href {\doibase
  10.1103/PhysRevD.110.034010} {\bibfield  {journal} {\bibinfo  {journal}
  {Phys. Rev. D}\ }\textbf {\bibinfo {volume} {110}},\ \bibinfo {pages}
  {034010} (\bibinfo {year} {2024})},\ \Eprint
  {http://arxiv.org/abs/2405.03105} {arXiv:2405.03105 [nucl-th]} \BibitemShut
  {NoStop}%
\bibitem [{\citenamefont {Daher}\ \emph
  {et~al.}(2024{\natexlab{b}})\citenamefont {Daher}, \citenamefont
  {Florkowski}, \citenamefont {Ryblewski},\ and\ \citenamefont
  {Taghinavaz}}]{Daher:2024bah}%
  \BibitemOpen
  \bibfield  {author} {\bibinfo {author} {\bibfnamefont {A.}~\bibnamefont
  {Daher}}, \bibinfo {author} {\bibfnamefont {W.}~\bibnamefont {Florkowski}},
  \bibinfo {author} {\bibfnamefont {R.}~\bibnamefont {Ryblewski}}, \ and\
  \bibinfo {author} {\bibfnamefont {F.}~\bibnamefont {Taghinavaz}},\ }\href
  {\doibase 10.1103/PhysRevD.109.114001} {\bibfield  {journal} {\bibinfo
  {journal} {Phys. Rev. D}\ }\textbf {\bibinfo {volume} {109}},\ \bibinfo
  {pages} {114001} (\bibinfo {year} {2024}{\natexlab{b}})},\ \Eprint
  {http://arxiv.org/abs/2403.04711} {arXiv:2403.04711 [hep-ph]} \BibitemShut
  {NoStop}%
\bibitem [{\citenamefont {Fang}\ \emph {et~al.}(2025)\citenamefont {Fang},
  \citenamefont {Fukushima}, \citenamefont {Pu},\ and\ \citenamefont
  {Wang}}]{Fang:2025aig}%
  \BibitemOpen
  \bibfield  {author} {\bibinfo {author} {\bibfnamefont {S.}~\bibnamefont
  {Fang}}, \bibinfo {author} {\bibfnamefont {K.}~\bibnamefont {Fukushima}},
  \bibinfo {author} {\bibfnamefont {S.}~\bibnamefont {Pu}}, \ and\ \bibinfo
  {author} {\bibfnamefont {D.-L.}\ \bibnamefont {Wang}},\ }\href@noop {} {\
  (\bibinfo {year} {2025})},\ \Eprint {http://arxiv.org/abs/2506.20698}
  {arXiv:2506.20698 [nucl-th]} \BibitemShut {NoStop}%
\bibitem [{\citenamefont {Montenegro}\ \emph {et~al.}(2017)\citenamefont
  {Montenegro}, \citenamefont {Tinti},\ and\ \citenamefont
  {Torrieri}}]{Montenegro:2017rbu}%
  \BibitemOpen
  \bibfield  {author} {\bibinfo {author} {\bibfnamefont {D.}~\bibnamefont
  {Montenegro}}, \bibinfo {author} {\bibfnamefont {L.}~\bibnamefont {Tinti}}, \
  and\ \bibinfo {author} {\bibfnamefont {G.}~\bibnamefont {Torrieri}},\ }\href
  {\doibase 10.1103/PhysRevD.96.056012} {\bibfield  {journal} {\bibinfo
  {journal} {Phys. Rev. D}\ }\textbf {\bibinfo {volume} {96}},\ \bibinfo
  {pages} {056012} (\bibinfo {year} {2017})},\ \bibinfo {note} {[Addendum:
  Phys.Rev.D 96, 079901 (2017)]},\ \Eprint {http://arxiv.org/abs/1701.08263}
  {arXiv:1701.08263 [hep-th]} \BibitemShut {NoStop}%
\bibitem [{\citenamefont {Montenegro}\ and\ \citenamefont
  {Torrieri}(2020)}]{Montenegro:2020paq}%
  \BibitemOpen
  \bibfield  {author} {\bibinfo {author} {\bibfnamefont {D.}~\bibnamefont
  {Montenegro}}\ and\ \bibinfo {author} {\bibfnamefont {G.}~\bibnamefont
  {Torrieri}},\ }\href {\doibase 10.1103/PhysRevD.102.036007} {\bibfield
  {journal} {\bibinfo  {journal} {Phys. Rev. D}\ }\textbf {\bibinfo {volume}
  {102}},\ \bibinfo {pages} {036007} (\bibinfo {year} {2020})},\ \Eprint
  {http://arxiv.org/abs/2004.10195} {arXiv:2004.10195 [hep-th]} \BibitemShut
  {NoStop}%
\bibitem [{\citenamefont {Abboud}\ \emph {et~al.}(2025)\citenamefont {Abboud},
  \citenamefont {Gavassino}, \citenamefont {Singh},\ and\ \citenamefont
  {Speranza}}]{Abboud:2025shb}%
  \BibitemOpen
  \bibfield  {author} {\bibinfo {author} {\bibfnamefont {N.}~\bibnamefont
  {Abboud}}, \bibinfo {author} {\bibfnamefont {L.}~\bibnamefont {Gavassino}},
  \bibinfo {author} {\bibfnamefont {R.}~\bibnamefont {Singh}}, \ and\ \bibinfo
  {author} {\bibfnamefont {E.}~\bibnamefont {Speranza}},\ }\href {\doibase
  10.1103/bngt-lbdv} {\bibfield  {journal} {\bibinfo  {journal} {Phys. Rev. D}\
  }\textbf {\bibinfo {volume} {112}},\ \bibinfo {pages} {094043} (\bibinfo
  {year} {2025})},\ \Eprint {http://arxiv.org/abs/2506.19786} {arXiv:2506.19786
  [nucl-th]} \BibitemShut {NoStop}%
\bibitem [{\citenamefont {Bhadury}\ \emph
  {et~al.}(2025{\natexlab{a}})\citenamefont {Bhadury}, \citenamefont {Drogosz},
  \citenamefont {Florkowski}, \citenamefont {Kar},\ and\ \citenamefont
  {Mykhaylova}}]{Bhadury:2025wuh}%
  \BibitemOpen
  \bibfield  {author} {\bibinfo {author} {\bibfnamefont {S.}~\bibnamefont
  {Bhadury}}, \bibinfo {author} {\bibfnamefont {Z.}~\bibnamefont {Drogosz}},
  \bibinfo {author} {\bibfnamefont {W.}~\bibnamefont {Florkowski}}, \bibinfo
  {author} {\bibfnamefont {S.~K.}\ \bibnamefont {Kar}}, \ and\ \bibinfo
  {author} {\bibfnamefont {V.}~\bibnamefont {Mykhaylova}},\ }\href@noop {} {\
  (\bibinfo {year} {2025}{\natexlab{a}})},\ \Eprint
  {http://arxiv.org/abs/2511.19295} {arXiv:2511.19295 [hep-ph]} \BibitemShut
  {NoStop}%
\bibitem [{\citenamefont {Florkowski}\ \emph
  {et~al.}(2019{\natexlab{a}})\citenamefont {Florkowski}, \citenamefont
  {Kumar},\ and\ \citenamefont {Ryblewski}}]{Florkowski:2018fap}%
  \BibitemOpen
  \bibfield  {author} {\bibinfo {author} {\bibfnamefont {W.}~\bibnamefont
  {Florkowski}}, \bibinfo {author} {\bibfnamefont {A.}~\bibnamefont {Kumar}}, \
  and\ \bibinfo {author} {\bibfnamefont {R.}~\bibnamefont {Ryblewski}},\ }\href
  {\doibase 10.1016/j.ppnp.2019.07.001} {\bibfield  {journal} {\bibinfo
  {journal} {Prog. Part. Nucl. Phys.}\ }\textbf {\bibinfo {volume} {108}},\
  \bibinfo {pages} {103709} (\bibinfo {year} {2019}{\natexlab{a}})},\ \Eprint
  {http://arxiv.org/abs/1811.04409} {arXiv:1811.04409 [nucl-th]} \BibitemShut
  {NoStop}%
\bibitem [{\citenamefont {Becattini}\ and\ \citenamefont
  {Lisa}(2020)}]{Becattini:2020ngo}%
  \BibitemOpen
  \bibfield  {author} {\bibinfo {author} {\bibfnamefont {F.}~\bibnamefont
  {Becattini}}\ and\ \bibinfo {author} {\bibfnamefont {M.~A.}\ \bibnamefont
  {Lisa}},\ }\href {\doibase 10.1146/annurev-nucl-021920-095245} {\bibfield
  {journal} {\bibinfo  {journal} {Ann. Rev. Nucl. Part. Sci.}\ }\textbf
  {\bibinfo {volume} {70}},\ \bibinfo {pages} {395} (\bibinfo {year} {2020})},\
  \Eprint {http://arxiv.org/abs/2003.03640} {arXiv:2003.03640 [nucl-ex]}
  \BibitemShut {NoStop}%
\bibitem [{\citenamefont {Huang}(2025)}]{Huang:2024ffg}%
  \BibitemOpen
  \bibfield  {author} {\bibinfo {author} {\bibfnamefont {X.-G.}\ \bibnamefont
  {Huang}},\ }\href {\doibase 10.1007/s41365-025-01784-3} {\bibfield  {journal}
  {\bibinfo  {journal} {Nucl. Sci. Tech.}\ }\textbf {\bibinfo {volume} {36}},\
  \bibinfo {pages} {208} (\bibinfo {year} {2025})},\ \Eprint
  {http://arxiv.org/abs/2411.11753} {arXiv:2411.11753 [nucl-th]} \BibitemShut
  {NoStop}%
\bibitem [{\citenamefont {Florkowski}(2025)}]{Florkowski:2024cif}%
  \BibitemOpen
  \bibfield  {author} {\bibinfo {author} {\bibfnamefont {W.}~\bibnamefont
  {Florkowski}},\ }\href {\doibase 10.1016/j.jspc.2025.100028} {\bibfield
  {journal} {\bibinfo  {journal} {J. Subatomic Part. Cosmol.}\ }\textbf
  {\bibinfo {volume} {3}},\ \bibinfo {pages} {100028} (\bibinfo {year}
  {2025})},\ \Eprint {http://arxiv.org/abs/2411.19673} {arXiv:2411.19673
  [hep-ph]} \BibitemShut {NoStop}%
\bibitem [{\citenamefont {Florkowski}\ and\ \citenamefont
  {Hontarenko}(2025)}]{Florkowski:2024bfw}%
  \BibitemOpen
  \bibfield  {author} {\bibinfo {author} {\bibfnamefont {W.}~\bibnamefont
  {Florkowski}}\ and\ \bibinfo {author} {\bibfnamefont {M.}~\bibnamefont
  {Hontarenko}},\ }\href {\doibase 10.1103/PhysRevLett.134.082302} {\bibfield
  {journal} {\bibinfo  {journal} {Phys. Rev. Lett.}\ }\textbf {\bibinfo
  {volume} {134}},\ \bibinfo {pages} {082302} (\bibinfo {year} {2025})},\
  \Eprint {http://arxiv.org/abs/2405.03263} {arXiv:2405.03263 [hep-ph]}
  \BibitemShut {NoStop}%
\bibitem [{\citenamefont {Drogosz}\ \emph {et~al.}(2024)\citenamefont
  {Drogosz}, \citenamefont {Florkowski},\ and\ \citenamefont
  {Hontarenko}}]{Drogosz:2024gzv}%
  \BibitemOpen
  \bibfield  {author} {\bibinfo {author} {\bibfnamefont {Z.}~\bibnamefont
  {Drogosz}}, \bibinfo {author} {\bibfnamefont {W.}~\bibnamefont {Florkowski}},
  \ and\ \bibinfo {author} {\bibfnamefont {M.}~\bibnamefont {Hontarenko}},\
  }\href {\doibase 10.1103/PhysRevD.110.096018} {\bibfield  {journal} {\bibinfo
   {journal} {Phys. Rev. D}\ }\textbf {\bibinfo {volume} {110}},\ \bibinfo
  {pages} {096018} (\bibinfo {year} {2024})},\ \Eprint
  {http://arxiv.org/abs/2408.03106} {arXiv:2408.03106 [hep-ph]} \BibitemShut
  {NoStop}%
\bibitem [{\citenamefont {Mathisson}(1937)}]{Mathisson:1937zz}%
  \BibitemOpen
  \bibfield  {author} {\bibinfo {author} {\bibfnamefont {M.}~\bibnamefont
  {Mathisson}},\ }\href
  {https://www.actaphys.uj.edu.pl/fulltext?series=T&vol=6&no=3&page=163}
  {\bibfield  {journal} {\bibinfo  {journal} {Acta Phys. Polon.}\ }\textbf
  {\bibinfo {volume} {6}},\ \bibinfo {pages} {163} (\bibinfo {year}
  {1937})}\BibitemShut {NoStop}%
\bibitem [{\citenamefont {{Mathisson}}(2010)}]{2010GReGr..42.1011M}%
  \BibitemOpen
  \bibfield  {author} {\bibinfo {author} {\bibfnamefont {M.}~\bibnamefont
  {{Mathisson}}},\ }\href {\doibase 10.1007/s10714-010-0939-y} {\bibfield
  {journal} {\bibinfo  {journal} {General Relativity and Gravitation}\ }\textbf
  {\bibinfo {volume} {42}},\ \bibinfo {pages} {1011} (\bibinfo {year}
  {2010})}\BibitemShut {NoStop}%
\bibitem [{\citenamefont {Bhadury}\ \emph
  {et~al.}(2025{\natexlab{b}})\citenamefont {Bhadury}, \citenamefont {Drogosz},
  \citenamefont {Florkowski}, \citenamefont {Kar},\ and\ \citenamefont
  {Mykhaylova}}]{Bhadury:2025boe}%
  \BibitemOpen
  \bibfield  {author} {\bibinfo {author} {\bibfnamefont {S.}~\bibnamefont
  {Bhadury}}, \bibinfo {author} {\bibfnamefont {Z.}~\bibnamefont {Drogosz}},
  \bibinfo {author} {\bibfnamefont {W.}~\bibnamefont {Florkowski}}, \bibinfo
  {author} {\bibfnamefont {S.~K.}\ \bibnamefont {Kar}}, \ and\ \bibinfo
  {author} {\bibfnamefont {V.}~\bibnamefont {Mykhaylova}},\ }\href@noop {} {\
  (\bibinfo {year} {2025}{\natexlab{b}})},\ \Eprint
  {http://arxiv.org/abs/2505.02657} {arXiv:2505.02657 [hep-ph]} \BibitemShut
  {NoStop}%
\bibitem [{\citenamefont {Drogosz}(2025)}]{Drogosz:2025ose}%
  \BibitemOpen
  \bibfield  {author} {\bibinfo {author} {\bibfnamefont {Z.}~\bibnamefont
  {Drogosz}},\ }\href {\doibase 10.3390/physics7030031} {\bibfield  {journal}
  {\bibinfo  {journal} {Physics}\ }\textbf {\bibinfo {volume} {7}},\ \bibinfo
  {pages} {31} (\bibinfo {year} {2025})},\ \Eprint
  {http://arxiv.org/abs/2504.03396} {arXiv:2504.03396 [hep-ph]} \BibitemShut
  {NoStop}%
\bibitem [{\citenamefont {Kar}\ and\ \citenamefont
  {Mykhaylova}(2025)}]{Kar:2025qvj}%
  \BibitemOpen
  \bibfield  {author} {\bibinfo {author} {\bibfnamefont {S.~K.}\ \bibnamefont
  {Kar}}\ and\ \bibinfo {author} {\bibfnamefont {V.}~\bibnamefont
  {Mykhaylova}},\ }\href@noop {} {\  (\bibinfo {year} {2025})},\ \Eprint
  {http://arxiv.org/abs/2511.09580} {arXiv:2511.09580 [quant-ph]} \BibitemShut
  {NoStop}%
\bibitem [{\citenamefont {Drogosz}(2026)}]{Drogosz:2025iyr}%
  \BibitemOpen
  \bibfield  {author} {\bibinfo {author} {\bibfnamefont {Z.}~\bibnamefont
  {Drogosz}},\ }\href {\doibase 10.1016/j.physletb.2026.140205} {\bibfield
  {journal} {\bibinfo  {journal} {Phys. Lett. B}\ }\textbf {\bibinfo {volume}
  {873}},\ \bibinfo {pages} {140205} (\bibinfo {year} {2026})},\ \Eprint
  {http://arxiv.org/abs/2509.06014} {arXiv:2509.06014 [hep-ph]} \BibitemShut
  {NoStop}%
\bibitem [{\citenamefont {Drogosz}\ \emph
  {et~al.}(2025{\natexlab{a}})\citenamefont {Drogosz}, \citenamefont
  {Florkowski},\ and\ \citenamefont {Mykhaylova}}]{Drogosz:2025ihp}%
  \BibitemOpen
  \bibfield  {author} {\bibinfo {author} {\bibfnamefont {Z.}~\bibnamefont
  {Drogosz}}, \bibinfo {author} {\bibfnamefont {W.}~\bibnamefont {Florkowski}},
  \ and\ \bibinfo {author} {\bibfnamefont {V.}~\bibnamefont {Mykhaylova}},\
  }\href {\doibase 10.1103/tg2w-czwq} {\bibfield  {journal} {\bibinfo
  {journal} {Phys. Rev. D}\ }\textbf {\bibinfo {volume} {112}},\ \bibinfo
  {pages} {L051901} (\bibinfo {year} {2025}{\natexlab{a}})},\ \Eprint
  {http://arxiv.org/abs/2506.01537} {arXiv:2506.01537 [hep-ph]} \BibitemShut
  {NoStop}%
\bibitem [{\citenamefont {Kreiss}\ \emph {et~al.}(1997)\citenamefont {Kreiss},
  \citenamefont {Nagy}, \citenamefont {Ortiz},\ and\ \citenamefont
  {Reula}}]{Kreiss:1997mk}%
  \BibitemOpen
  \bibfield  {author} {\bibinfo {author} {\bibfnamefont {H.-O.}\ \bibnamefont
  {Kreiss}}, \bibinfo {author} {\bibfnamefont {G.~B.}\ \bibnamefont {Nagy}},
  \bibinfo {author} {\bibfnamefont {O.~E.}\ \bibnamefont {Ortiz}}, \ and\
  \bibinfo {author} {\bibfnamefont {O.~A.}\ \bibnamefont {Reula}},\ }\href
  {\doibase 10.1063/1.531940} {\bibfield  {journal} {\bibinfo  {journal} {J.
  Math. Phys.}\ }\textbf {\bibinfo {volume} {38}},\ \bibinfo {pages} {5272}
  (\bibinfo {year} {1997})},\ \Eprint {http://arxiv.org/abs/gr-qc/9702008}
  {arXiv:gr-qc/9702008} \BibitemShut {NoStop}%
\bibitem [{\citenamefont {Florkowski}\ \emph
  {et~al.}(2019{\natexlab{b}})\citenamefont {Florkowski}, \citenamefont
  {Kumar}, \citenamefont {Ryblewski},\ and\ \citenamefont
  {Singh}}]{Florkowski:2019qdp}%
  \BibitemOpen
  \bibfield  {author} {\bibinfo {author} {\bibfnamefont {W.}~\bibnamefont
  {Florkowski}}, \bibinfo {author} {\bibfnamefont {A.}~\bibnamefont {Kumar}},
  \bibinfo {author} {\bibfnamefont {R.}~\bibnamefont {Ryblewski}}, \ and\
  \bibinfo {author} {\bibfnamefont {R.}~\bibnamefont {Singh}},\ }\href
  {\doibase 10.1103/PhysRevC.99.044910} {\bibfield  {journal} {\bibinfo
  {journal} {Phys. Rev. C}\ }\textbf {\bibinfo {volume} {99}},\ \bibinfo
  {pages} {044910} (\bibinfo {year} {2019}{\natexlab{b}})},\ \Eprint
  {http://arxiv.org/abs/1901.09655} {arXiv:1901.09655 [hep-ph]} \BibitemShut
  {NoStop}%
\bibitem [{\citenamefont {Drogosz}\ \emph
  {et~al.}(2025{\natexlab{b}})\citenamefont {Drogosz}, \citenamefont
  {Florkowski}, \citenamefont {\L{}ygan},\ and\ \citenamefont
  {Ryblewski}}]{Drogosz:2024lkx}%
  \BibitemOpen
  \bibfield  {author} {\bibinfo {author} {\bibfnamefont {Z.}~\bibnamefont
  {Drogosz}}, \bibinfo {author} {\bibfnamefont {W.}~\bibnamefont {Florkowski}},
  \bibinfo {author} {\bibfnamefont {N.}~\bibnamefont {\L{}ygan}}, \ and\
  \bibinfo {author} {\bibfnamefont {R.}~\bibnamefont {Ryblewski}},\ }\href
  {\doibase 10.1103/PhysRevC.111.024909} {\bibfield  {journal} {\bibinfo
  {journal} {Phys. Rev. C}\ }\textbf {\bibinfo {volume} {111}},\ \bibinfo
  {pages} {024909} (\bibinfo {year} {2025}{\natexlab{b}})},\ \Eprint
  {http://arxiv.org/abs/2411.06154} {arXiv:2411.06154 [hep-ph]} \BibitemShut
  {NoStop}%
\bibitem [{\citenamefont {Singh}\ \emph {et~al.}(2025)\citenamefont {Singh},
  \citenamefont {Ryblewski},\ and\ \citenamefont {Florkowski}}]{Singh:2024cub}%
  \BibitemOpen
  \bibfield  {author} {\bibinfo {author} {\bibfnamefont {S.~K.}\ \bibnamefont
  {Singh}}, \bibinfo {author} {\bibfnamefont {R.}~\bibnamefont {Ryblewski}}, \
  and\ \bibinfo {author} {\bibfnamefont {W.}~\bibnamefont {Florkowski}},\
  }\href {\doibase 10.1103/PhysRevC.111.024907} {\bibfield  {journal} {\bibinfo
   {journal} {Phys. Rev. C}\ }\textbf {\bibinfo {volume} {111}},\ \bibinfo
  {pages} {024907} (\bibinfo {year} {2025})},\ \Eprint
  {http://arxiv.org/abs/2411.08223} {arXiv:2411.08223 [hep-ph]} \BibitemShut
  {NoStop}%
\bibitem [{\citenamefont {Sapna}\ \emph {et~al.}(2025)\citenamefont {Sapna},
  \citenamefont {Singh},\ and\ \citenamefont {Wagner}}]{Sapna:2025yss}%
  \BibitemOpen
  \bibfield  {author} {\bibinfo {author} {\bibnamefont {Sapna}}, \bibinfo
  {author} {\bibfnamefont {S.~K.}\ \bibnamefont {Singh}}, \ and\ \bibinfo
  {author} {\bibfnamefont {D.}~\bibnamefont {Wagner}},\ }\href {\doibase
  10.1103/1s6g-fs8w} {\bibfield  {journal} {\bibinfo  {journal} {Phys. Rev. C}\
  }\textbf {\bibinfo {volume} {112}},\ \bibinfo {pages} {054902} (\bibinfo
  {year} {2025})},\ \Eprint {http://arxiv.org/abs/2503.22552} {arXiv:2503.22552
  [hep-ph]} \BibitemShut {NoStop}%
\bibitem [{\citenamefont {Bjorken}(1983)}]{Bjorken:1982qr}%
  \BibitemOpen
  \bibfield  {author} {\bibinfo {author} {\bibfnamefont {J.~D.}\ \bibnamefont
  {Bjorken}},\ }\href {\doibase 10.1103/PhysRevD.27.140} {\bibfield  {journal}
  {\bibinfo  {journal} {Phys. Rev. D}\ }\textbf {\bibinfo {volume} {27}},\
  \bibinfo {pages} {140} (\bibinfo {year} {1983})}\BibitemShut {NoStop}%
\bibitem [{\citenamefont {Florkowski}(2010)}]{Florkowski:2010zz}%
  \BibitemOpen
  \bibfield  {author} {\bibinfo {author} {\bibfnamefont {W.}~\bibnamefont
  {Florkowski}},\ }\href@noop {} {\emph {\bibinfo {title} {{Phenomenology of
  Ultra-Relativistic Heavy-Ion Collisions}}}}\ (\bibinfo  {publisher}
  {Singapore: World Scientific},\ \bibinfo {year} {2010})\BibitemShut {NoStop}%
%%CITATION = INSPIRE-868078;%%
\bibitem [{\citenamefont {Coleman}(2018)}]{Coleman:2018mew}%
  \BibitemOpen
  \bibfield  {author} {\bibinfo {author} {\bibfnamefont {S.}~\bibnamefont
  {Coleman}},\ }\href {\doibase 10.1142/9371} {\emph {\bibinfo {title}
  {{Lectures of Sidney Coleman on Quantum Field Theory}}}},\ edited by\
  \bibinfo {editor} {\bibfnamefont {B.~G.-g.}\ \bibnamefont {Chen}}, \bibinfo
  {editor} {\bibfnamefont {D.}~\bibnamefont {Derbes}}, \bibinfo {editor}
  {\bibfnamefont {D.}~\bibnamefont {Griffiths}}, \bibinfo {editor}
  {\bibfnamefont {B.}~\bibnamefont {Hill}}, \bibinfo {editor} {\bibfnamefont
  {R.}~\bibnamefont {Sohn}}, \ and\ \bibinfo {editor} {\bibfnamefont {Y.-S.}\
  \bibnamefont {Ting}}\ (\bibinfo  {publisher} {WSP},\ \bibinfo {address}
  {Hackensack},\ \bibinfo {year} {2018})\BibitemShut {NoStop}%
\bibitem [{\citenamefont {Kapusta}\ \emph
  {et~al.}(2020{\natexlab{a}})\citenamefont {Kapusta}, \citenamefont {Rrapaj},\
  and\ \citenamefont {Rudaz}}]{Kapusta:2019sad}%
  \BibitemOpen
  \bibfield  {author} {\bibinfo {author} {\bibfnamefont {J.~I.}\ \bibnamefont
  {Kapusta}}, \bibinfo {author} {\bibfnamefont {E.}~\bibnamefont {Rrapaj}}, \
  and\ \bibinfo {author} {\bibfnamefont {S.}~\bibnamefont {Rudaz}},\ }\href
  {\doibase 10.1103/PhysRevC.101.024907} {\bibfield  {journal} {\bibinfo
  {journal} {Phys. Rev. C}\ }\textbf {\bibinfo {volume} {101}},\ \bibinfo
  {pages} {024907} (\bibinfo {year} {2020}{\natexlab{a}})},\ \Eprint
  {http://arxiv.org/abs/1907.10750} {arXiv:1907.10750 [nucl-th]} \BibitemShut
  {NoStop}%
\bibitem [{\citenamefont {Kapusta}\ \emph
  {et~al.}(2020{\natexlab{b}})\citenamefont {Kapusta}, \citenamefont {Rrapaj},\
  and\ \citenamefont {Rudaz}}]{Kapusta:2020npk}%
  \BibitemOpen
  \bibfield  {author} {\bibinfo {author} {\bibfnamefont {J.~I.}\ \bibnamefont
  {Kapusta}}, \bibinfo {author} {\bibfnamefont {E.}~\bibnamefont {Rrapaj}}, \
  and\ \bibinfo {author} {\bibfnamefont {S.}~\bibnamefont {Rudaz}},\ }\href
  {\doibase 10.1103/PhysRevC.102.064911} {\bibfield  {journal} {\bibinfo
  {journal} {Phys. Rev. C}\ }\textbf {\bibinfo {volume} {102}},\ \bibinfo
  {pages} {064911} (\bibinfo {year} {2020}{\natexlab{b}})},\ \Eprint
  {http://arxiv.org/abs/2004.14807} {arXiv:2004.14807 [hep-th]} \BibitemShut
  {NoStop}%
\bibitem [{\citenamefont {Florkowski}\ \emph
  {et~al.}(2018{\natexlab{b}})\citenamefont {Florkowski}, \citenamefont
  {Friman}, \citenamefont {Jaiswal}, \citenamefont {Ryblewski},\ and\
  \citenamefont {Speranza}}]{Florkowski:2017dyn}%
  \BibitemOpen
  \bibfield  {author} {\bibinfo {author} {\bibfnamefont {W.}~\bibnamefont
  {Florkowski}}, \bibinfo {author} {\bibfnamefont {B.}~\bibnamefont {Friman}},
  \bibinfo {author} {\bibfnamefont {A.}~\bibnamefont {Jaiswal}}, \bibinfo
  {author} {\bibfnamefont {R.}~\bibnamefont {Ryblewski}}, \ and\ \bibinfo
  {author} {\bibfnamefont {E.}~\bibnamefont {Speranza}},\ }\href {\doibase
  10.1103/PhysRevD.97.116017} {\bibfield  {journal} {\bibinfo  {journal} {Phys.
  Rev. D}\ }\textbf {\bibinfo {volume} {97}},\ \bibinfo {pages} {116017}
  (\bibinfo {year} {2018}{\natexlab{b}})},\ \Eprint
  {http://arxiv.org/abs/1712.07676} {arXiv:1712.07676 [nucl-th]} \BibitemShut
  {NoStop}%
\bibitem [{\citenamefont {Montes}\ \emph {et~al.}(2023)\citenamefont {Montes},
  \citenamefont {Rubio},\ and\ \citenamefont {Reula}}]{Montes:2023dex}%
  \BibitemOpen
  \bibfield  {author} {\bibinfo {author} {\bibfnamefont {P.~E.}\ \bibnamefont
  {Montes}}, \bibinfo {author} {\bibfnamefont {M.~E.}\ \bibnamefont {Rubio}}, \
  and\ \bibinfo {author} {\bibfnamefont {O.~A.}\ \bibnamefont {Reula}},\ }\href
  {\doibase 10.1103/PhysRevD.107.103041} {\bibfield  {journal} {\bibinfo
  {journal} {Phys. Rev. D}\ }\textbf {\bibinfo {volume} {107}},\ \bibinfo
  {pages} {103041} (\bibinfo {year} {2023})},\ \Eprint
  {http://arxiv.org/abs/2304.08584} {arXiv:2304.08584 [gr-qc]} \BibitemShut
  {NoStop}%
\bibitem [{\citenamefont {Bemfica}(2025)}]{Bemfica:2025gws}%
  \BibitemOpen
  \bibfield  {author} {\bibinfo {author} {\bibfnamefont {F.~S.}\ \bibnamefont
  {Bemfica}},\ }\href {\doibase 10.1103/blhw-xplr} {\bibfield  {journal}
  {\bibinfo  {journal} {Phys. Rev. E}\ }\textbf {\bibinfo {volume} {112}},\
  \bibinfo {pages} {065105} (\bibinfo {year} {2025})},\ \Eprint
  {http://arxiv.org/abs/2508.04717} {arXiv:2508.04717 [physics.gen-ph]}
  \BibitemShut {NoStop}%
\bibitem [{\citenamefont {Keeble}\ and\ \citenamefont
  {Pretorius}(2025)}]{Keeble:2025bkc}%
  \BibitemOpen
  \bibfield  {author} {\bibinfo {author} {\bibfnamefont {L.~S.}\ \bibnamefont
  {Keeble}}\ and\ \bibinfo {author} {\bibfnamefont {F.}~\bibnamefont
  {Pretorius}},\ }\href {\doibase 10.1103/d4wd-zj7w} {\bibfield  {journal}
  {\bibinfo  {journal} {Phys. Rev. D}\ }\textbf {\bibinfo {volume} {112}},\
  \bibinfo {pages} {124034} (\bibinfo {year} {2025})},\ \Eprint
  {http://arxiv.org/abs/2508.20998} {arXiv:2508.20998 [gr-qc]} \BibitemShut
  {NoStop}%
\end{thebibliography}

\end{document}